\documentclass[aps,prb,preprint,superscriptaddress,longbibliography]{revtex4-2}
\usepackage{hyperref}
\usepackage{epsfig}
\usepackage{graphicx}
\usepackage{subfigure}
\usepackage{latexsym}
\usepackage{color}
\usepackage{fullpage}
\usepackage{amssymb}
\usepackage{dcolumn}
\usepackage{bm}
\usepackage[normalem]{ulem}
\usepackage{units}
\usepackage{amsmath}
\usepackage[paperwidth=210mm,paperheight=297mm,centering,hmargin=2cm,vmargin=2.3cm]{geometry}
\usepackage{eqnarray,amsmath}
\usepackage{float}
\usepackage{soul}

\def\xsigmaLambda#1{\includegraphics[width=#1em]{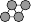}}
\def\sigmaLambda{{
  \mathchoice
    {\xsigmaLambda1}%
    {\xsigmaLambda1}%
    {\xsigmaLambda\defaultscriptratio}%
    {\xsigmaLambda\defaultscriptscriptratio}}}

\def\xsigmaGamma#1{\includegraphics[width=#1em]{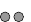}}
\def\sigmaGamma{{
  \mathchoice
    {\xsigmaGamma1}%
    {\xsigmaGamma1}%
    {\xsigmaGamma\defaultscriptratio}%
    {\xsigmaGamma\defaultscriptscriptratio}}}

\def\xthreeUp#1{\includegraphics[width=#1em]{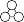}}
\def\threeUp{{
  \mathchoice
    {\xthreeUp1}%
    {\xthreeUp1}%
    {\xthreeUp\defaultscriptratio}%
    {\xthreeUp\defaultscriptscriptratio}}}

\def\xthreeDown#1{\includegraphics[width=#1em]{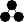}}
\def\threeDown{{
  \mathchoice
    {\xthreeDown1}%
    {\xthreeDown1}%
    {\xthreeDown\defaultscriptratio}%
    {\xthreeDown\defaultscriptscriptratio}}}

\def\xtwoUp#1{\includegraphics[width=#1em]{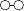}}
\def\twoUp{{
  \mathchoice
    {\xtwoUp1}%
    {\xtwoUp1}%
    {\xtwoUp\defaultscriptratio}%
    {\xtwoUp\defaultscriptscriptratio}}}

\def\xtwoDown#1{\includegraphics[width=#1em]{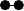}}
\def\twoDown{{
  \mathchoice
    {\xtwoDown1}%
    {\xtwoDown1}%
    {\xtwoDown\defaultscriptratio}%
    {\xtwoDown\defaultscriptscriptratio}}}

\setcitestyle{super} 

\begin{document}



\title{Probing magnetic correlations in space and time within predefined topological sectors of a macroscopic spin liquid}

\author{R\'{e}my Dangoisse }
\author{Jeanne Colbois}
\author{Laurent Del Rey}
\author{Nicolas Rougemaille}
\email{nicolas.rougemaille@neel.cnrs.fr}
\author{Johann Coraux}
\email{johann.coraux@neel.cnrs.fr}
\affiliation{Universit\'{e} Grenoble Alpes, CNRS, Institut NEEL, Grenoble INP, 38000 Grenoble, France}

\begin{abstract}

The triangular Ising antiferromagnet, with its residual entropy density and critical correlations at absolute zero, is the archetype of a two-dimensional spin liquid. Its ground state is partitioned into topological sectors connected by global spin flip events that wrap the lattice boundaries, and become statistically irrelevant in the thermodynamic limit where magnetic fluctuations are restricted to the dominant sector. Thus far, these properties have been mainly investigated from a theoretical perspective, and one might wonder to what extent they can be transposed to real materials. Here, we present experimental observations obtained in an artificial, macroscopic realisation of the seminal triangular Ising antiferromagnet that consists of a lattice of millimeter-sized NdFeB cylinders put into motion by a mechanical shaker. Specifically, we demonstrate that the very-low-energy physics and the true ground state of this model can be reached experimentally. Besides, we are able to probe, in space and time, the magnetic properties within manually preselected pockets of the ground-state manifold that emulate, to a good approximation, the behaviour in distinct topological sectors. Our approach opens new avenues for naked-eye visualisation and hand manipulation of many-body phenomena associated to frustrated magnetism and models of statistical physics.

\end{abstract}

\maketitle

\newpage

\section*{Introduction}

Onsager showed that a second-order phase transition brings a two-dimensional (2D) Ising ferromagnet from a high-temperature paramagnet to a low-temperature ordered phase \cite{Onsager1944}. This brought the far-reaching idea that magnetic order emerges as a cooperative behaviour, even in systems governed by short range spin interactions. Soon after, the case of an Ising antiferromagnet on a 2D triangular lattice was considered by Wannier \cite{Wannier1950}. There, antiferromagnetism does not fit in, to borrow his words, and neither magnetic ordering nor a phase transition occur, demonstrating that correlated disorder too can emerge as a cooperative behaviour in a short range spin model \cite{Wannier1950,Houtappel1950,Husimi1950,Newell1950,Temperley1950}. Although conceptually simple, this model exhibits a rich palette of intriguing properties: its ground state is extensively degenerate, i.e., has a nonzero entropy density at low temperature, is critical at 0~K \cite{Stephenson1964}, exhibits a configuration space partitioned into topological sectors \cite{Smerald2016}, and allows several powerful mappings with models of statistical mechanics.

However influential this archetypal model of frustrated magnetism has been, its framework is restrictive: in actual crystals, spin layers are rarely purely 2D, the degree of freedom may not be truly of Ising type, the lattice often deviates from a perfect triangular one, and interactions generally extend beyond nearest neighbours. The singular spin liquid physics of the triangular Ising antiferromagnet (TIAF) model hence remains elusive experimentally, whether one considers ferroic crystals \cite{Collins1997,Little2020,Bastien2024,Zhu2025}, 2D alloys \cite{Ottaviano2003,Azizi2020} or molecular monolayers \cite{Charra1998,AlfonsoMoro2023}. After all, and Onsager asked the same about his own model, is the TIAF too simple and far removed from natural crystals? The question seems fair, and for the time being physicists are still left with the option of designing artificial systems to bypass what Nature provides \cite{Davidovic1997,Wang2006,Nisoli2013,Tierno2019,Rougemaille2019,Skjaervo2020}, thereby emulating different kinds of on-lattice Ising models, frustrated \cite{Tanaka2005,Qi2008,Rougemaille2011,Zhang2013,Anghinolfi2015,Perrin2016,Ostman2018,Schanilec2022} or not \cite{Arnalds2016,Nguyen2017,Ostman2018b}, and the behaviour of their excitations \cite{Farhan2019,May2021}. Maybe surprisingly, only a few of these man-made frustrated triangular lattices \cite{Han2008,Farhan2020,Pip2021,Pac2025,Wang2025} address the physics of the TIAF model, and when they do, they usually face severe limitations to reach the true ground state manifold of the critical spin liquid, in which all cooperative phenomena emerge. For example, arrays of interacting nanomagnets \cite{Pip2021,Pac2025} or millimeter-sized magnets \cite{Wang2025} leave the system barely correlated within the paramagnetic regime, in some cases with excitation densities exceeding 20\% \cite{Wang2025}.

In this work, we have designed an experimental platform at the macroscopic scale that does emulate the intriguing physics of the celebrated TIAF ground state. As we will see below, this platform consists of permanent magnets that can slide within mm-sized cavities upon mechanical shaking (Fig.~\ref{fig:continuousIsing}). When the shaking protocol is properly adjusted, our setup allows direct visualisation of how the lattices dynamically approach a spin liquid regime from an arbitrary defined initial configuration. In particular, starting from a magnetically saturated configuration, i.e., the highest energy state for the TIAF model, we monitor how magnetic correlations develop in space as a function of time, as the system equilibrates. Surprisingly, our measurements reveal that the finite-size nature of our lattices leads to a two-step correlation mechanism as it reaches the ground state: strong spin-spin correlations first set in within the bulk of the system, giving rise to a spin liquid, before an antiferromagnetic ordering of the spins develops all along the lattice periphery. In other words, spin order and spin disorder coexist in the ground state because of the finite size of the system, resulting in a reduced, yet extensive, residual entropy density as compared to the one derived by Wannier \cite{Wannier1950,Wannier1973}. This key observation then guides us to specifically manipulate, by hand, some of the lattice spins---those at the edge. When they are fixed in an antiferromagnetic fashion, the system efficiently and reproducibly reaches ground state configurations of the TIAF model. More interestingly, the spin arrangement along the lattice edges can be arbitrary defined, for example to emulate simili periodic boundary conditions, or to force the system to fluctuate within a predefined topological sector. The unique capabilities of our platform to reach the low-energy manifold of the TIAF model and to permit full control over the boundary conditions, provide access to many-body properties that could not be envisioned so far.

\section*{Results}

\subsection*{A macroscopic mechanical emulator of the TIAF model}

\begin{figure}[!hbt]
\begin{center}
\includegraphics[width=80mm]{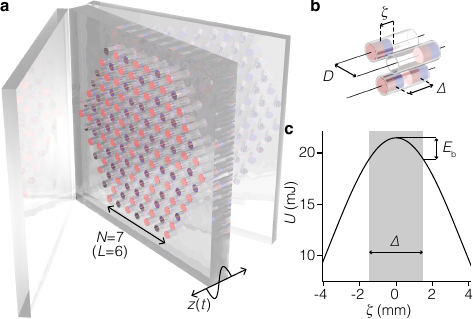} 
\caption{\label{fig:continuousIsing}Macroscopic TIAF emulator. (a) Schematics of the triangular lattice of cavities, filled with NdFeB cylinders, and driven in sinusoidal motion $z(t)$ by a mechanical shaker. Lattices with $N=7$ and 11 magnets per edge have been manufactured. (b) Three neighbour magnets; $D=8.7$~mm ($N=7$) or $D=6.0$~mm ($N=11$) is the lattice period and $\zeta \in[-\Delta/2,+\Delta/2]$, the magnet position, with $\Delta=3$~mm, which defines the $\pm$1 value of an Ising degree of freedom. (c) Interaction potential $U$ vs $\zeta$; an energy barrier $E_\mathrm{b}$ separates the two degenerate low-energy states.} 
\end{center}
\end{figure}

Our system consists of mm-sized NdFeB cylinders that can slide along their long axis, within closed cavities drilled inside a plexiglass plate (Fig.~\ref{fig:continuousIsing}a and Methods), similar to what has been introduced recently \cite{Wang2025,Ge2025}. The cylinders are manually inserted in the cavities, one by one, in such a way that they are in mutual repulsive magnetostatic interaction (the north and south poles of the magnets are all oriented in the same direction, see red/blue colors in Fig.~\ref{fig:continuousIsing}a and Fig.~\ref{fig:continuousIsing}b). The cavities being longer than the magnets they host, two neighboring cylinders tend to increase their separating distance (see Fig.~\ref{fig:continuousIsing}b), such that in a lattice each magnet can take only one of two possible positions, thus altogether behaving as Ising spins similar to what happens in artificial colloidal spin ices \cite{Tierno2016}. The lattices are hexagon-shaped to stick with the symmetry of the triangular lattice and to probe magnetic fluctuations within predefined topological sectors as we will see below. In the following, we consider $N$=7 and $N$=11 magnets per edge, i.e., 127 and 331 spins, respectively, as schematised in Fig.~\ref{fig:continuousIsing}a. The transparent plexiglass plate allows observation of the magnets positions using a standard camera. Our experimental setup then falls into the category of macroscopic artificial platforms that give access to naked-eye visualization of one-dimensional (1D) and 2D phenomena \cite{Olive1998,Kirschner2003,Mellado2012,Velo2020,Goncalves2020,Teixeira2024,Peroor2025,Scafuri2025}.

Maximising the magnet-to-magnet distance requires crossing an energy barrier, which we estimated analytically (Fig.~\ref{fig:continuousIsing}c) and experimentally through force measurements (about 2~mJ, see Supplementary Note~1). Hence, making the system fluctuates requires a nonzero energy input.  This energy is supplied by a mechanical shaker that forces the lattice to vibrate horizontally (Supplementary Fig.~2) at tunable amplitude and frequency. 

\begin{figure*}[!hbt] 
\begin{center}
\includegraphics[width=124.36mm]{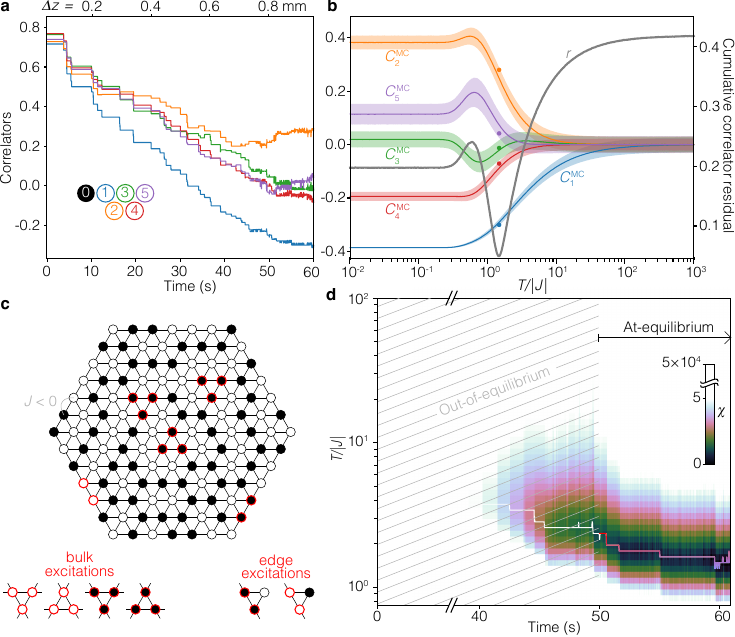}
\caption{\label{fig:dynamics}Naked eye visualisation of equilibration and concerted spin dynamics. (a) Evolution of the first five spin correlators $C_{k\leq5}$ during a stepwise demagnetization, starting from an almost saturated configuration, and as the shaking amplitude $\Delta z$ increases stepwise over time (step duration: 0.1~s). (b) $C^{\mathrm{MC}}_{k\leq5}$ as function of temperature from Monte Carlo simulations (shaded areas represent the standard deviation), experimental data points at 60~s (colored dots) and cumulative correlator residual $r(T)$. (c) Experimental configuration obtained at 60~s, with all bulk $\threeUp$ / $\threeDown$ excitations, and a few of the edge $\twoUp$ / $\twoDown$ excitations highlighted. (d) Reduced cumulative correlator residuals $\chi(T)$ as function of time. The temperature minimising $\chi$ is overlaid, and becomes an equilibrium temperature when $\chi<2$.} 
\end{center}
\end{figure*}

\subsection*{Reaching low-energy configurations from an arbitrarily defined initial state}

To illustrate the operation of our experimental platform, we consider in the following an initial state wherein (almost all, 85\%) the magnets have been placed the same way in the lattice cavities, i.e., corresponding to a magnetically saturated configuration. Ramping up the shaking oscillation amplitude $\Delta z$ triggers the switching of an increasing number of magnets, eventually producing a disordered mixture of up and down magnets. To analyse this disorder and to quantify how it sets in, we examined the correlations in the magnets' positions in space and time. In the language of magnetism, this corresponds to computing the spin-spin correlators defined as $C_k = \langle s_i \cdot s_j \rangle = \frac{1}{n} \sum_{i=1}^n \frac{1}{n_k} \sum_{j=1}^{n_k} s_i \cdot s_j$, with $n_k$ the number of spins in the $k^{\mathrm{th}}$ shell around an Ising spin $s_i$ residing on site $i$ of the lattice ($C_{k>5}$ are averaged over only a small fraction of the spins, less than 30\% due to the finite size of the lattice, thus have relatively weak statistical significance, which is why we discard them).  Figure~\ref{fig:dynamics}a shows how the $C_{k\leq5}$ values change with time, as the shaking amplitude increases. The staircase evolution of the measured correlations are associated to the reversal of individual (or collections of) spins, whereas the plateaus reveal that the lattice remains frozen as long as $\Delta z$ is not further increased. As expected, the flipping rate increases with $\Delta z$ (compare this rate for $\Delta z$ = 0.2 and 0.8~mm for instance). Besides, the strong positive values of the close-to-saturated initial state decrease as it could be anticipated. However, the five correlators do not necessarily tend to zero at the end of the shaking protocol, demonstrating that the system ends up into a correlated disorder regime. Notably, the second and fifth correlator values increase after reaching a minimum value around $\Delta z$ = 0.8~mm, suggesting that the shaking amplitude might allow the system to equilibrate.

Comparing the final experimental $C_{k}$ values to those computed with Monte Carlo simulations, $C^{\mathrm{MC}}_{k}$ (see Methods), provides a powerful and well-established mean \cite{Chioar2014,Chioar2014b,Hofhuis2020,Pac2025} to determine to what extent the observed configurations are representative of the at-equilibrium physics of a model Ising spin Hamiltonian (Supplementary Note~2). Figure~\ref{fig:dynamics}b reports the temperature dependence of the $C^{\mathrm{MC}}_{k\leq5}$ correlators obtained numerically, and reveals that the measurements can be accurately described by the TIAF model at an effective temperature $T$ of about the strength of coupling constant ($|J|$, here $J<0$) between nearest neighbours (see how the colored dots in Fig.~\ref{fig:dynamics}b are well captured by the Monte Carlo simulations). Minimising the cumulative correlator residual $r(T)=\sqrt{\sum_{k\leq 5} [C^{\mathrm{MC}}_{k}(T) - C_{k}]^2}$ indeed returns an effective temperature of 1.35$\times |J|$ (Supplementary Note~3), showing that our lattice starts to correlate, while keeping a substantial amount of excitations. These excitations are of two kinds (Fig.~\ref{fig:dynamics}c): away from the lattice edges, they are triplets of magnets having the same height ($\threeUp$, $\threeDown$), at the edges, they are pairs of magnets with the same height ($\twoUp$, $\twoDown$). Their presence affects the $C_{1}$ correlator (when the system is fully correlated, i.e., when local excitations are exponentially suppressed, $C_{1}$ should reach the $-1/3$ value). We note that this effective temperature, although still high, is already below those reported in arrays of interacting nanomagnets \cite{Pip2021,Pac2025}.

Interestingly, a good fit with Monte Carlo simulations is obtained not only at 60~s (when the shaking is turned off), but before that, over a $\sim$10~s time-span (Fig.~\ref{fig:dynamics}a). There, we can define an effective equilibrium temperature, albeit slightly higher than at 60~s, as the spin correlations dynamically evolve (Supplementary Fig.~3). However, such a temperature cannot be defined throughout the entire time sequence. This is illustrated in Fig.~\ref{fig:dynamics}d, where the reduced cumulative correlator residuals $\chi(T)$, i.e., $r(T)$ normalized to the simulations' statistical dispersion, are stacked (horizontally) over time. Satisfactorily small values of $\chi(T)$'s minimum, of the order of unity, are only attained after 50~s (dark green region in Fig.~\ref{fig:dynamics}c). At earlier times, the system must be considered out-of-equilibrium (see an alternative representation in Supplementary Note~5). The capability of our experimental platform to capture the spatial and temporal evolutions of the magnetic correlations provides a powerful mean to track how a low-energy configuration of the TIAF model is reached starting from a given initial state.

\begin{figure*}[!hbt] 
\begin{center}
\includegraphics[width=160mm]{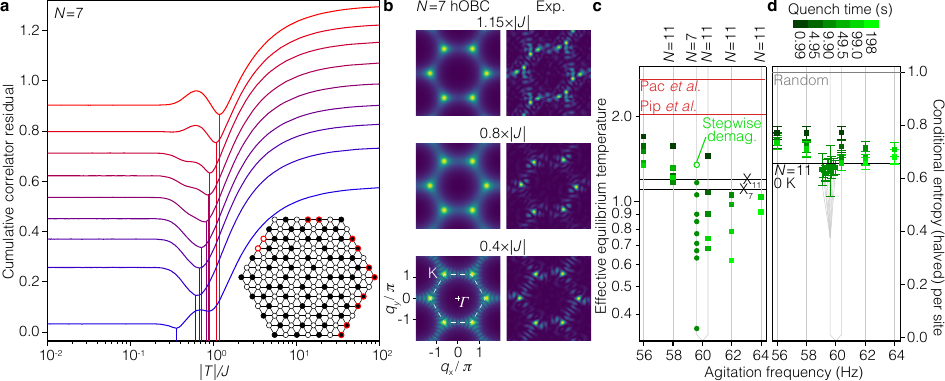}
\caption{\label{fig:lowT}Low and tunable equilibrium temperatures. (a) Cumulative residual of the five first correlators relative to Monte Carlo simulations (OBC), as function of temperature, using a driving protocol. The 40 experimental configurations have been grouped in nine collections of configurations, and their average $C_{k\leq5}$ correlators each produce a curve. Curves have been vertically shifted, each by 0.1, for clarity. The curves' minima define an effective equilibrium temperature. The configuration shown as inset corresponds to the lowest (effective) equilibrium temperature. (b) Magnetic structure factors at different (effective) temperatures, from the Monte Carlo and experimental data. (c) Summary of effective equilibrium temperatures (standard deviations are smaller than the symbol size) for lattices of sizes $N=7$ and $N=11$ using the driving protocol, varying the quench time and agitation frequency (for $N=11$); the effective equilibrium temperature reached for $N=7$ via stepwise demagnetisation (as in Fig.~\ref{fig:dynamics}b), and the value obtained with arrays of nanomagnets \cite{Pip2021,Pac2025} (red lines) are also shown. X$_7$ and X$_{11}$ mark the crossover temperatures for the two lattice sizes. (d) Conditional entropy (with its experimental standard deviation) corresponding to the temperatures in (c). Expectations for the TIAF's (PBC, $N=11$) ground state and a randomly disordered configuration are shown as horizontal lines.} 
\end{center}
\end{figure*}

\subsection*{Approaching the ground state: a two-step process}

At this point, we have shown that our shaking protocol brings the system to an equilibrated configuration, characterized by a reasonably low effective temperature of the order of the nearest-neighbour coupling strength. We might now wonder whether this effective temperature can be further reduced, and if so, whether this can be achieved in a reproducible manner. As we will see below, the answer is yes to both questions.

Increasing the shaking amplitude $\Delta z$ above the highest value reported in Fig.~\ref{fig:dynamics}a leads to an interesting observation. When this amplitude exceeds typically a millimeter, the motion of the magnetic cylinders is driven by the oscillation of the plexiglass plate, i.e., the displacement of the magnets is synchronized with that of the support for a finite time during each period of the oscillation. As a consequence of their inertia, the cylinders hit the ends of their cavities repeatedly, at the oscillation frequency. This motion, which can be partly monitored using stroboscopic imaging (see Supplementary Note~6), presumably provides enough energy to efficiently overcome the barrier separating the two Ising states. The resulting configuration is then imaged after the shaking amplitude has been ramped down to zero within a minute, typically. This shaking protocol, different from the one addressed in Fig.~\ref{fig:dynamics}, first drives the spin configuration of the entire lattice before allowing it to fall into an energy minimum as the power is slowly turned off. It is, in a sense, reminiscent of the field demagnetization protocols employed in the case of athermal nanomagnetic arrays \cite{Wang2007,Nisoli2007,Ke2008,Rougemaille2011,Morgan2013}.

This shaking protocol has been applied a hundred times on our $N=7$ lattice. Successively-generated configurations are, to a large extent, decorrelated from one another, with all magnets in the lattice having changed states at least a few times over the full set of configurations, although these changes do not appear to occur fully uniformly over the lattice (see Supplementary Fig.~7 and Supplementary Note~7 about possible sources of inhomogeneity). Among the set of 100 configurations, those with $\min(r)$ greater than 10\% are disregarded. For such configurations, a description in terms of a thermal TIAF ensemble indeed is hardly relevant as not all spin-spin correlators, sometimes none of them, match (within the statistical dispersion) those predicted by the Monte Carlo simulations (Supplementary Fig.~8)---they are thus considered out-of-equilibrium. The 40 configurations with $\min(r)<10\%$ have been grouped in nine subsets of three to six configurations showing similar $C_{k\leq5}$ values, and the $r(T)$ analysis reported in Fig.~\ref{fig:dynamics}b has been repeated each time to determine the average effective equilibrium temperature of the subsets. Figure~\ref{fig:lowT}a shows that these temperatures range from 0.36 to 1.17$\times |J|$, i.e., unprecedentedly small values, due to the vanishing density of $\threeUp$ and $\threeDown$ excitations (0, 1 or 2 per lattice, see Supplementary Note~8). We should stress that if the shaking protocol brings the system to low effective temperatures, it is not particularly reproducible at this point since only 40\% of the thus-produced configurations are well described by a thermal TIAF ensemble. We will show later how this moderate reproducibility can be improved considerably.

If the correlation analysis allows us to estimate the effective temperature associated to the measured magnetic configuration, computing the magnetic structure factor (MSF) generally provides complementary pieces of information. The MSFs were thus derived from the experimental configurations and compared to those calculated from our Monte Carlo simulations (see Methods). Results, reported in Fig.~\ref{fig:lowT}b, reveal good agreement between the experimental and numerical MSFs, as could be anticipated by the fact that all considered configurations are representative of (at-equilibrium) TIAF model (according to our analysis of the first five spin correlators). At first sight, the MSFs do not show substantial temperature dependence, and they exhibit strongest intensity at the K points of the Brillouin zone, consistent with the antiferromagnetic nature of the nearest-neighbour correlations \cite{Smerald2016,AlfonsoMoro2023}. However, closer inspection shows that the MSFs' diffuse intensity bridging the Brillouin zone's corners (found in nanomagnets arrays implementing the TIAF \cite{Pac2025} or an analog of it \cite{Farhan2020}) becomes structured for low enough (effective) temperature, in the spin liquid regime. This is clearly apparent for the experimental configurations, further illustrating the agreement (here qualitative) in the correlation data (here in reciprocal space) between the experimental and Monte Carlo simulation data found in real space ($C_{k\leq5}$) analysis in Fig.~\ref{fig:lowT}a and Supplementary Fig.~8a. The structuration at the Brillouin zone edges is even more striking in the numerical MSFs, owing to the large number of configurations provided by the Monte Carlo simulations (10,000 per temperature), from which well-sampled averages are computed (experimental MSFs are computed from single configurations). At the lowest temperature in particular, the intensity exhibits clear oscillations. These features along the edges of the Brillouin zones signal magnetic correlations evolving substantially as the low-energy manifold is approached. Going back to real space, the spin configurations at the lowest effective temperatures indeed reveal that the spins at the lattice periphery have a tendency to be ordered and anti-aligned (see inset of Fig.~\ref{fig:lowT}a). In other words, if most of the lattice appears disordered and liquid-like, its edges ultimately order in an antiferromagnetic fashion. In hindsight, some properties of the ground state can be understood from a simple splitting of the Hamiltonian into bulk and boundary contributions (see Supplementary Note~9). Our correlation analysis thus reveals that the TIAF model under open, hexagonal-shaped boundary conditions correlates in a two-step process: liquid-like correlations already develop before the lattice periphery orders. Said differently, the density of $\threeUp$ and $\threeDown$ bulk excitations first vanishes as the temperature decreases, before the density of $\twoUp$ and $\twoDown$ edge excitations is reduced.

We note that at the lowest effective temperature ($0.36 \times |J|$), a nonzero density of $\twoUp$ and $\twoDown$ excitations is still found at the lattice edges. These edge excitations are in fact difficult to eliminate with a single spin-flip dynamics as they are constrained by the bulk, excitation-free configuration. Their elimination at no cost requires multiple single spin flips occurring in very specific sequences, overall representing very low-probability many-body space- and time-correlated recombinations (see Supplementary Fig.~9). We also note that the number of oscillations visible in the MSF intensity along the Brillouin zone edges is directly linked to the lattice size. In the measurements reported here, seven maxima in the intensity can be identified, which corresponds to the $N=7$ lattice size. Increasing $N$ accordingly leads to an increased number of intensity maxima (see Supplementary Fig.~10 for the case $N=11$), unambiguously demonstrating that the MSF intensity along the Brillouin zones' edges is associated to the ordered spin configuration at the lattice edges.

\subsection*{Residual entropy}

\begin{figure}[!hbt]
\begin{center}
\includegraphics[width=74.13mm]{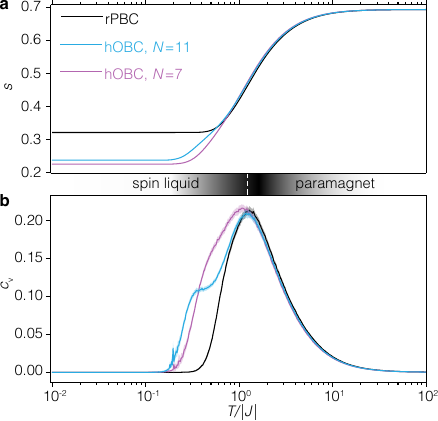} 
\caption{\label{fig:thermo}Thermodynamics of the TIAF under different boundary conditions. (a) Entropy density ($s$) and (b) specific heat ($c_\mathrm{v}$, together with its standard deviation shown as a shaded area) as function of temperature for a rhombus-shaped TIAF in periodic boundary conditions (rPBC, 121 spins) and hexagon-shaped lattices (two sizes, $N$ = 7, 11, so 127 and 331 spins) in open boundary conditions (hOBC), according to Monte Carlo simulations (see Methods).} 
\end{center}
\end{figure}

Knowing that the lattice edges get ordered at low temperature, we might wonder how much the entropy density is reduced compared to the usual case with period boundary conditions. To answer this question, we have computed the thermodynamic properties of the TIAF model for $N=7$ and $N=11$, both with open (OBC) and periodic (PBC) boundary conditions. As anticipated, the residual entropy is reduced with OBC, but importantly, it is nonzero and still large, in agreement with our previous observation that the system remains a spin liquid in the bulk (Fig.~\ref{fig:thermo}a). With no surprise, the reduction of the entropy density $s$ is larger for small lattice sizes
(see Supplementary Note~9). If the two-step correlation mechanism manifests itself through an entropy reduction, it also shows up as a shoulder or a double bump in the specific heat $c_\mathrm{v}$ (Fig.~\ref{fig:thermo}b). The main bump is associated to the usual paramagnetic $\rightarrow$ spin liquid crossover, whereas the second one at lower temperature is only present for finite-size lattices and moves to lower temperatures when the lattice size increases. We hence assign it to the ordering of the spins at the lattice edges. The numerical simulations thus are fully consistent with our experimental findings.

In magnetic bulk compounds, determining the residual entropy experimentally is relatively standard, once the ensemble-averaged temperature-dependent $c_{\mathrm{v}}$ has been measured. However, the latter is very imprecisely evaluated with most artificial systems. Besides, the practical impossibility to measure $c_{\mathrm{v}}$ over a broad temperature range precludes an estimate of $s$. Nevertheless, an estimate for the entropy (upper bound in the infinite size limit) can be given using conditional-probability arguments, as demonstrated in artificial arrays of nanomagnets \cite{Lammert2010}. Here, we use the following bound estimate \cite{Farhan2020} $s\leq S(\,\sigmaLambda\,|\,\sigmaGamma)/2$, where $S(\,\sigmaLambda\,|\,\sigmaGamma)$ is the conditional entropy (see Supplementary Note~12).

The $S(\,\sigmaLambda\,|\,\sigmaGamma)$ values obtained with the $N=7$ lattice are characteristic of a correlated disorder, and are significantly lower than in the case of random disorder. No clear correlation emerges between $S(\,\sigmaLambda\,|\,\sigmaGamma)$ and the effective equilibrium temperatures, presumably due to the small lattice size (Fig.~\ref{fig:lowT}c and Fig.~\ref{fig:lowT}d). On the contrary, for $N=11$, the configurations with lowest effective temperature (0.62 $\times |J|$) are indeed those having the lowest entropy (Fig.~\ref{fig:lowT}c and Fig.~\ref{fig:lowT}d), and the range of variation of the conditional entropy nicely matches the estimate derived from Monte Carlo configurations (see Supplementary Note~12). From an experimental perspective, we note that an optimum is reached in the shaking protocol for a certain frequency and quench time (i.e., the time needed to ramp the power of the shaker down to zero once the configuration of the lattice is fully driven by the mechanical excitation). Once those conditions have been identified, and this is a central result of our work, we are able to guide our mechanical TIAF emulator at an effective temperature below the crossover bringing the system from the paramagnetic regime to the spin liquid state (the crossover temperatures, assessed from the Monte Carlo simulations, are marked with $X_7$ and $X_{11}$ for both lattice sizes in Fig.~\ref{fig:lowT}c), which has so far remained inaccessible \cite{Ottaviano2003,Azizi2020,Pip2021,Pac2025} (Supplementary Note~13).

\subsection*{True ground state, emulation of periodic boundary conditions and fluctuations within predefined topological sectors via boundary conditions engineering}

\begin{figure*}[!hbt]
\begin{center}
\includegraphics[width=107.84mm]{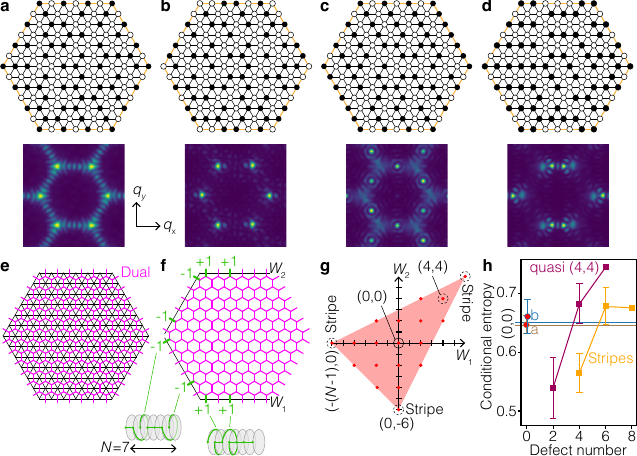}
\caption{\label{fig:boundaries}Boundary condition engineering and ground state topological sectors. (a-c) Experimental configurations with fixed (a) alternated, (b) disordered, (c) mixed alternated+all-down and (d) mixed alternated+disorder arrangements of spins at the lattice edges (along the orange lines). All configurations comply with PBC. Note the absence of excitations in (a,b). Below each configuration, MSF averaged over 32 (a,b), 20 (c) and 40 (d) different configurations is shown. (e) Triangular lattice's dual, a honeycomb lattice (pink). (f) Dimer mapping for the configuration in (b), with a green dimer placed on the segments of the dual lattice that are surrounded by $\twoUp$ or $\twoDown$ doublets. Dimers away from the lattice edges are not represented. Each dimer is associated to a winding, with positive or negative sign depending on the lattice edge, and the winding numbers are sums over two paths, defining the topological numbers $W_1$ and $W_2$. (g) Allowed pairs (red points) of topological winding numbers ($W_1$,$W_2$) in the TIAF's ground state (in PBC). The configurations shown in (a,b) both correspond to the (0,0) topological sector. The configurations in (c,d) ressemble configurations in the stripe quasi (-6,0) and (4,4) sectors respectively; however they comprise $\threeUp$ and $\threeDown$ excitations and thus do not belong to the ground state. (h) Conditional entropy (and its experimental standard deviation) in the (a-d) kinds of configurations as function of the number of excitation in the configurations, compared to the expectations for the (0,0) ground state (horizontal lines, corresponding to edge configurations as in (a), in brown, and in (b), in blue).} 
\end{center}
\end{figure*}

We have seen above that reproducibly obtaining configurations free of excitations, especially at the lattice edges, i.e. achieving perfect antiferromagnetic spin ordering there as in the true ground state of the TIAF, is challenging with our shaking protocol. What we do next is to enforce this alternation by fixing the outer spin states manually (Fig.~\ref{fig:boundaries}a), using nonmagnetic rods blocking the motion of the corresponding magnets---a strategy that can only be transposed to few other kinds of artificial platforms \cite{Rodriguez2021,Baillou2026}. Repeating the same protocol now leads to a collection of configurations, a large fraction (65\%) of which are free of $\threeUp$ and $\threeDown$ excitations, exhibit correlated disorder away from the lattice edges (Fig.~\ref{fig:boundaries}a), and are indiscernible from those of the true TIAF ground state (obtained with our Monte Carlo simulations). Imposing the boundary conditions thus allows us to circumvent the main experimental limitation associated to the annihilation of $\twoUp$ or $\twoDown$ edge excitations. This way, the macroscopic mechanical system emulates the TIAF model's true ground state with a fairly high probability, with no excitation present in 32 configurations such as in Fig.~\ref{fig:boundaries}a, generated experimentally. This is confirmed by comparing the experimental and numerical MSFs which now become hardly distinguishable (Fig.~\ref{fig:lowT}b and Fig.~\ref{fig:boundaries}a).

At this point, it is instructive to realise that fixing the spin alternation at the lattice periphery, we have made all facing edges have the very same configuration, as if the lattice were in PBC. Although by doing so we cannot sample over all edge configurations as in true PBC, the system can somehow be seen as wrapped on a torus, in which one line of spins in each direction is fixed (see Supplementary Fig.~13d). Besides forcing ground state configurations experimentally, engineering the boundary conditions at will offers the appealing opportunity to emulate, at least to some extent, magnetic fluctuations within a preselected topological sector. Indeed, the TIAF model is known to have a ground state manifold that is partitioned into sectors, i.e., pockets in the configuration space that are connected one another by global loop moves wrapping the system boundaries \cite{Nienhuis1984,Millane2004,Smerald2016,Smerald2018}. In other words, once the system enters the ground state manifold, the configurations then belong to a well-defined topological sector: the single-spin-flip dynamics and local loop moves make the system fluctuates within this topological sector. The other topological sectors remain statistically impossible to visit, unless global loop moves wrapping the lattice boundaries are permitted. In practice, ergodicity is broken if one only considers single-spin-flip dynamics, and the system is trapped, forever, within a given topological sector. Imposing the boundary conditions manually allows us to predefine the topological sector (provided that the systems exhibits no excitation, as in the TIAF ground state), in which the system may fluctuate.

The system's topology can be apprehended using a mapping of the spin configuration onto a dimer one, living on the dual of the triangular lattice (the honeycomb lattice, see Fig.~\ref{fig:boundaries}e) \cite{Kasteleyn1963,Smerald2016,Smerald2018}. In this representation, dimers are then placed halfway between $\twoUp$ and $\twoDown$ doublets (green segments in Fig.~\ref{fig:boundaries}f). Counting the number of dimers crossing the edge lines defines two winding numbers $W_1$ and $W_2$ (with a particular sign convention, see Fig.~\ref{fig:boundaries}f,g). Configurations with other $(W_1,W_2)$ values belong to one of the 19 different topological sectors \cite{Smerald2016,Smerald2018} allowed for a lattice with $L=N-1=6$ maximum number of dimers at the lattice edges (Fig.~\ref{fig:boundaries}g). 

We have then prepared different edge configurations with distinct winding numbers and we have applied the shaking protocol to reach configurations free of $\threeUp$ and $\threeDown$ excitations (edge $\twoUp$ and $\twoDown$ excitations are now irrelevant). We note that the $(0,0)$ sector dominates the ground state manifold of the TIAF model, totaling $>$99\% of the possible configurations at the thermodynamic limit \cite{Smerald2016}. Within this sector, spin-alternated-edge configurations such as the one shown in Fig.~\ref{fig:boundaries}a constitute a peculiar, yet extensively degenerate subset of the $(0,0)$ sector. Figure~\ref{fig:boundaries}b represents another subset. There, two dimers cross each edge, and no obvious order is apparent along the lattice periphery. The MSF computed over 32 experimental configurations having the same (fixed) edge configuration (Fig.~\ref{fig:boundaries}b) does not show the intensity oscillations along the edges of the Brillouin zone that are observed in the first case (Fig.~\ref{fig:boundaries}a), demonstrating that spin correlations differ within subsets of the $(0,0)$ sector. Our experimental platform allows us to probe directly how correlations develop within the main topological sector.

Using the same approach, we also have the option to probe other sectors, including those that are statistically irrelevant at the thermodynamic limit. For example, the (-6,0), (0,-6) and (6,6) sectors correspond to ground state configurations exhibiting a stripe pattern. This is illustrated in (Fig.~\ref{fig:boundaries}c), where stripe correlations show up in the MSF as peaks at the M point of the Brillouin zone, despite the presence of a few remaining $\threeUp$ and $\threeDown$ excitations. We could argue that such excitations preclude the description of the manifold in terms of topological sectors; in more rigorous words, what we do here is emulating a topological-sector-like subset, rather than predefining an actual topological sector. Nevertheless, we can still describe such configurations using what may be called quasi winding numbers, which are however not topological invariants anymore (see Supplementary Note~14 for a more comprehensive framework based on three, and not just two, quasi winding numbers). As another example, the (4,4) quasi winding numbers have also be implemented  (Fig.~\ref{fig:boundaries}d). The configurations obtained with our shaking protocol there again comprise a small but nonzero amount of $\threeUp$ and $\threeDown$ excitations. The averaged MSF shows yet another kind of pattern, reflecting the lattice edges' nontrivial spin arrangement, and also featuring diffuse contributions related to the spin disorder in the lattice.

We end this section by considering the conditional entropy of the boundary-engineered configurations. Both the edge-alternated and edge-disordered excitation-free collections of (0,0) configurations have nonzero conditional entropy (Fig.~\ref{fig:boundaries}h)---a ground state property. Our stripes and quasi-(4,4) configurations have nonzero entropy too, but with larger values when the defect density increases. Conversely, as the system's effective temperature approaches that of the TIAF ground state, with fewer and fewer defects, its entropy reduces strongly, more prominently for the stripe configurations (compare the burgundy and orange points for the case of 4 defects), which is indeed the least degenerate (only threefold).

\subsection*{Prospects}

The amenability of our artificial system to dynamical imaging and local spin manipulations, combined to considerable freedom for lattice design, offer unique perspectives, for example to explore slow (glassy) dynamics in dipolar kagome magnets \cite{Hamp2018,Cugliandolo2020}, or nonconventional phase transitions such as the Kasteleyn one or a Devil's staircase physics \cite{Fisher1980,Smerald2018,Rufino2025}. Our experimental platform also offers new prospects for investigating different kinds of mappings. We shortly addressed one of them (a mapping to dimers), but others exist, in particular a mapping to strings \cite{Yokoi1986,Jiang2006,Kenyon2009}, which will bring a correspondance between the 2D classical problem (TIAF) and a lower-dimensional, 1D quantum model \cite{Samuel1980,Yokoi1986,Jiang2006,Smerald2018}. There, strings may be understood as the classical counterpart of worldlines of free spinless fermions within chains and evolving in imaginary time. This viewpoint could be investigated in the future, to devise an appropriate emerging theory, wherein the energy densities of the fermions and the free energy of the strings are equivalent, provided that the fermion mass is mapped to the string tension, and $\hbar$ to a virtual temperature accounting for the strings' entropic fluctuations \cite{Jiang2006}.

\newpage

\section*{Methods}

\subsection*{Lattice fabrication, mechanical shaking and data acquisition}

Lattices of cavities were engineered through plain plexiglass plates with accurate (within 5~$\mu$m) automated drilling. Lattices of different pitch were drilled, to find the optimal range of inter-magnet interactions (here restricted to first neighbours). The cavities are 2.1~mm in diameter for a hexagon-shaped lattice with $N=7$ magnets per edge, 1.15~mm for $N=11$ lattices. Magnetic cylinders (2~mm diameter and 3~mm height for $N=7$, 1~mm and 2~mm for $N=11$), made of nickel-coated NdFeB, were purchased from Supermagnete, and filled manually, one-by-one, inside the plexiglass plate.

A SignalForce GW-V20 mechanical shaker, fed with an air-cooled PA300E power supply, was used to agitate the lattices, by periodic translation $z(t)$ at frequency $f$ and with amplitude $\Delta z$, along the horizontal direction (which is also the axis of the cavities wherein the magnets slide), without rotation. The lattices were held in vertical position with four few-millimeter-wide (square cross-section) plexiglass beams (Supplementary Fig.~2). The power output of the shaker was controlled via an input sinusoidal bias provided by a Tektronix AFG1000 device, itself controlled via sets of command line instructions launched within the pyvisa library as implemented within a python 3.8.10 environment.

Photographs and videos were acquired with a Lumix G90 Panasonic camera. High frame-rate (60~fps), high-resolution data (1280$\times$720 pixels) were transfered in real-time via a DeckLink Mini Recorder 4K capture video card. The statistical analysis of the data sets (thousand-frame videos for data as shown in Fig.~\ref{fig:dynamics} or several thousands of configurations generated to produce the data in Figs.~\ref{fig:lowT},\ref{fig:boundaries}) prevents man-made analysis of the lattices and requires instead an automated code, which was written in python 3.8.10. A convolutional neural network (as implemented in the pyTorch library) has been used for that purpose, and trained over about 10,000 lattice sites (with magnets in either of their Ising state, so showing a different optical contrast). This approach was combined with a user-defined automatic detection of the planar lattice geometry.

\subsection*{Magnetic structure factor calculations}

Knowing the set of positions of each spin in a lattice (of $n$ sites labeled by $i$, $j$ indices) and their spin state ($s_{i,j}$), the magnetic structure factor is calculated as

\begin{equation}
S(\bm{q})=\frac{1}{n}\sum_{i,j} s_i s_j \times e^{i \bm{q}\cdot\bm{r}_{ij}},
\label{eq:MSF}
\end{equation}

\noindent with $\bm{r}_{ij}$ the vector connecting two lattice sites. More efficiently, $S(\bm{q})$ is computed using

\begin{equation}
S(\bm{q})=\frac{1}{n}||\bm{v}_q||^2,
\label{eq:MSFbis}
\end{equation}

\noindent with $\bm{v}_q=\sum_{i} \sigma_i e^{i \bm{q}\cdot\bm{r}_i}$. 
In practice, the $S(\bm{q})$ are averaged over several 10 experimental configurations and 10,000 configurations generated by Monte Carlo simulations. The calculation is made using home-made python and C++ multithreaded codes.

\subsection*{Monte Carlo simulations}

Monte Carlo simulations using the Metropolis algorithm were implemented within two home-made python and C++ multithreaded codes. The thermodynamics were computed with 100,000 configurations, with a 1.02 ratio between succesive temperatures, using 5,000 or 10,000 thermalization steps; the steps actually being modified Monte Carlo steps correcting for the reduction of the acceptance rate as temperature is decreased. For Fig.~\ref{fig:thermo} three kinds of lattices have been considered: one with a rhombus shape and 11 spins per edge, so 121 spins in total, in periodic boundary conditions, and two others one with a hexagon shape, with $N=7$ spins per edge, so 127 spins in total, and with $N=11$ (so 331 spins) in open boundary conditions (as in the experiments).

\bigskip

\section*{Data availability}

The numerical data supporting this study are available via Zenodo at \url{https://doi.org/10.5281/zenodo.17659176}. 
Source data are provided with this paper.


\begin{thebibliography}{73}%
  \makeatletter
  \providecommand \@ifxundefined [1]{%
   \@ifx{#1\undefined}
  }%
  \providecommand \@ifnum [1]{%
   \ifnum #1\expandafter \@firstoftwo
   \else \expandafter \@secondoftwo
   \fi
  }%
  \providecommand \@ifx [1]{%
   \ifx #1\expandafter \@firstoftwo
   \else \expandafter \@secondoftwo
   \fi
  }%
  \providecommand \natexlab [1]{#1}%
  \providecommand \enquote  [1]{``#1''}%
  \providecommand \bibnamefont  [1]{#1}%
  \providecommand \bibfnamefont [1]{#1}%
  \providecommand \citenamefont [1]{#1}%
  \providecommand \href@noop [0]{\@secondoftwo}%
  \providecommand \href [0]{\begingroup \@sanitize@url \@href}%
  \providecommand \@href[1]{\@@startlink{#1}\@@href}%
  \providecommand \@@href[1]{\endgroup#1\@@endlink}%
  \providecommand \@sanitize@url [0]{\catcode `\\12\catcode `\$12\catcode
    `\&12\catcode `\#12\catcode `\^12\catcode `\_12\catcode `\%12\relax}%
  \providecommand \@@startlink[1]{}%
  \providecommand \@@endlink[0]{}%
  \providecommand \url  [0]{\begingroup\@sanitize@url \@url }%
  \providecommand \@url [1]{\endgroup\@href {#1}{\urlprefix }}%
  \providecommand \urlprefix  [0]{URL }%
  \providecommand \Eprint [0]{\href }%
  \providecommand \doibase [0]{https://doi.org/}%
  \providecommand \selectlanguage [0]{\@gobble}%
  \providecommand \bibinfo  [0]{\@secondoftwo}%
  \providecommand \bibfield  [0]{\@secondoftwo}%
  \providecommand \translation [1]{[#1]}%
  \providecommand \BibitemOpen [0]{}%
  \providecommand \bibitemStop [0]{}%
  \providecommand \bibitemNoStop [0]{.\EOS\space}%
  \providecommand \EOS [0]{\spacefactor3000\relax}%
  \providecommand \BibitemShut  [1]{\csname bibitem#1\endcsname}%
  \let\auto@bib@innerbib\@empty
  \bibitem [{\citenamefont {Onsager}(1944)}]{Onsager1944}%
    \BibitemOpen
    \bibfield  {author} {\bibinfo {author} {\bibfnamefont {L.}~\bibnamefont
    {Onsager}},\ }\bibfield  {title} {\bibinfo {title} {Crystal statistics. {I}.
    a two-dimensional model with an order-disorder transition},\ }\href
    {https://doi.org/10.1103/PhysRev.65.117} {\bibfield  {journal} {\bibinfo
    {journal} {Phys. Rev.}\ }\textbf {\bibinfo {volume} {65}},\ \bibinfo {pages}
    {117} (\bibinfo {year} {1944})}\BibitemShut {NoStop}%
  \bibitem [{\citenamefont {Wannier}(1950)}]{Wannier1950}%
    \BibitemOpen
    \bibfield  {author} {\bibinfo {author} {\bibfnamefont {G.~H.}\ \bibnamefont
    {Wannier}},\ }\bibfield  {title} {\bibinfo {title} {Antiferromagnetism. the
    triangular {Ising} net},\ }\href {https://doi.org/10.1103/PhysRev.79.357}
    {\bibfield  {journal} {\bibinfo  {journal} {Phys. Rev.}\ }\textbf {\bibinfo
    {volume} {79}},\ \bibinfo {pages} {357} (\bibinfo {year} {1950})}\BibitemShut
    {NoStop}%
  \bibitem [{\citenamefont {Houtappel}(1950)}]{Houtappel1950}%
    \BibitemOpen
    \bibfield  {author} {\bibinfo {author} {\bibfnamefont {R.}~\bibnamefont
    {Houtappel}},\ }\bibfield  {title} {\bibinfo {title} {Order-disorder in
    hexagonal lattices},\ }\href
    {https://doi.org/https://doi.org/10.1016/0031-8914(50)90130-3} {\bibfield
    {journal} {\bibinfo  {journal} {Physica}\ }\textbf {\bibinfo {volume} {16}},\
    \bibinfo {pages} {425} (\bibinfo {year} {1950})}\BibitemShut {NoStop}%
  \bibitem [{\citenamefont {Husimi}\ and\ \citenamefont
    {Sy{\^o}zi}(1950)}]{Husimi1950}%
    \BibitemOpen
    \bibfield  {author} {\bibinfo {author} {\bibfnamefont {K.}~\bibnamefont
    {Husimi}}\ and\ \bibinfo {author} {\bibfnamefont {I.}~\bibnamefont
    {Sy{\^o}zi}},\ }\bibfield  {title} {\bibinfo {title} {The statistics of
    honeycomb and triangular lattice. {I}},\ }\href
    {https://doi.org/10.1143/ptp/5.2.177} {\bibfield  {journal} {\bibinfo
    {journal} {Progr. Theor. Phys.}\ }\textbf {\bibinfo {volume} {5}},\ \bibinfo
    {pages} {177} (\bibinfo {year} {1950})}\BibitemShut {NoStop}%
  \bibitem [{\citenamefont {Newell}(1950)}]{Newell1950}%
    \BibitemOpen
    \bibfield  {author} {\bibinfo {author} {\bibfnamefont {G.~F.}\ \bibnamefont
    {Newell}},\ }\bibfield  {title} {\bibinfo {title} {Crystal statistics of a
    two-dimensional triangular {Ising} lattice},\ }\href
    {https://doi.org/10.1103/PhysRev.79.876} {\bibfield  {journal} {\bibinfo
    {journal} {Phys. Rev.}\ }\textbf {\bibinfo {volume} {79}},\ \bibinfo {pages}
    {876} (\bibinfo {year} {1950})}\BibitemShut {NoStop}%
  \bibitem [{\citenamefont {Temperley}(1950)}]{Temperley1950}%
    \BibitemOpen
    \bibfield  {author} {\bibinfo {author} {\bibfnamefont {H.~N.~V.}\
    \bibnamefont {Temperley}},\ }\bibfield  {title} {\bibinfo {title}
    {Statistical mechanics of the two-dimensional assembly},\ }\href
    {https://doi.org/10.1098/rspa.1950.0094} {\bibfield  {journal} {\bibinfo
    {journal} {Proc. R. Soc. Lond. A}\ }\textbf {\bibinfo {volume} {202}},\
    \bibinfo {pages} {202} (\bibinfo {year} {1950})}\BibitemShut {NoStop}%
  \bibitem [{\citenamefont {Stephenson}(1964)}]{Stephenson1964}%
    \BibitemOpen
    \bibfield  {author} {\bibinfo {author} {\bibfnamefont {J.}~\bibnamefont
    {Stephenson}},\ }\bibfield  {title} {\bibinfo {title} {Ising‐model spin
    correlations on the triangular lattice},\ }\href
    {https://doi.org/10.1063/1.1704202} {\bibfield  {journal} {\bibinfo
    {journal} {J. Math. Phys.}\ }\textbf {\bibinfo {volume} {5}},\ \bibinfo
    {pages} {1009} (\bibinfo {year} {1964})}\BibitemShut {NoStop}%
  \bibitem [{\citenamefont {Smerald}\ \emph {et~al.}(2016)\citenamefont
    {Smerald}, \citenamefont {Korshunov},\ and\ \citenamefont
    {Mila}}]{Smerald2016}%
    \BibitemOpen
    \bibfield  {author} {\bibinfo {author} {\bibfnamefont {A.}~\bibnamefont
    {Smerald}}, \bibinfo {author} {\bibfnamefont {S.}~\bibnamefont {Korshunov}},\
    and\ \bibinfo {author} {\bibfnamefont {F.}~\bibnamefont {Mila}},\ }\bibfield
    {title} {\bibinfo {title} {Topological aspects of symmetry breaking in
    triangular-lattice {Ising} antiferromagnets},\ }\href
    {https://doi.org/10.1103/PhysRevLett.116.197201} {\bibfield  {journal}
    {\bibinfo  {journal} {Phys. Rev. Lett.}\ }\textbf {\bibinfo {volume} {116}},\
    \bibinfo {pages} {197201} (\bibinfo {year} {2016})}\BibitemShut {NoStop}%
  \bibitem [{\citenamefont {Collins}\ and\ \citenamefont
    {Petrenko}(1997)}]{Collins1997}%
    \BibitemOpen
    \bibfield  {author} {\bibinfo {author} {\bibfnamefont {M.~F.}\ \bibnamefont
    {Collins}}\ and\ \bibinfo {author} {\bibfnamefont {O.~A.}\ \bibnamefont
    {Petrenko}},\ }\bibfield  {title} {\bibinfo {title} {Review/synth\`{e}se:
    Triangular antiferromagnets},\ }\href {https://doi.org/10.1139/p97-007}
    {\bibfield  {journal} {\bibinfo  {journal} {Can. J. Phys.}\ }\textbf
    {\bibinfo {volume} {75}},\ \bibinfo {pages} {605} (\bibinfo {year}
    {1997})}\BibitemShut {NoStop}%
  \bibitem [{\citenamefont {Little}\ \emph {et~al.}(2020)\citenamefont {Little},
    \citenamefont {Lee}, \citenamefont {John}, \citenamefont {Doyle},
    \citenamefont {Maniv}, \citenamefont {Nair}, \citenamefont {Chen},
    \citenamefont {Rees}, \citenamefont {Venderbos}, \citenamefont {Fernandes},
    \citenamefont {Analytis},\ and\ \citenamefont {Orenstein}}]{Little2020}%
    \BibitemOpen
    \bibfield  {author} {\bibinfo {author} {\bibfnamefont {A.}~\bibnamefont
    {Little}}, \bibinfo {author} {\bibfnamefont {C.}~\bibnamefont {Lee}},
    \bibinfo {author} {\bibfnamefont {C.}~\bibnamefont {John}}, \bibinfo {author}
    {\bibfnamefont {S.}~\bibnamefont {Doyle}}, \bibinfo {author} {\bibfnamefont
    {E.}~\bibnamefont {Maniv}}, \bibinfo {author} {\bibfnamefont {N.~L.}\
    \bibnamefont {Nair}}, \bibinfo {author} {\bibfnamefont {W.}~\bibnamefont
    {Chen}}, \bibinfo {author} {\bibfnamefont {D.}~\bibnamefont {Rees}}, \bibinfo
    {author} {\bibfnamefont {J.~W.}\ \bibnamefont {Venderbos}}, \bibinfo {author}
    {\bibfnamefont {R.~M.}\ \bibnamefont {Fernandes}}, \bibinfo {author}
    {\bibfnamefont {J.~G.}\ \bibnamefont {Analytis}},\ and\ \bibinfo {author}
    {\bibfnamefont {J.}~\bibnamefont {Orenstein}},\ }\bibfield  {title} {\bibinfo
    {title} {Three-state nematicity in the triangular lattice antiferromagnet
    {Fe$_{1/3}$NbS$_2$}},\ }\href {https://doi.org/10.1038/s41563-020-0681-0}
    {\bibfield  {journal} {\bibinfo  {journal} {Nat. Mater.}\ }\textbf {\bibinfo
    {volume} {19}},\ \bibinfo {pages} {1062} (\bibinfo {year}
    {2020})}\BibitemShut {NoStop}%
  \bibitem [{\citenamefont {Bastien}\ \emph {et~al.}(2024)\citenamefont
    {Bastien}, \citenamefont {Rep{\v{c}}ek}, \citenamefont {Eli{\'a}{\v{s}}},
    \citenamefont {Kancko}, \citenamefont {Courtade}, \citenamefont {Haidamak},
    \citenamefont {Savinov}, \citenamefont {Bovtun}, \citenamefont {Kempa},
    \citenamefont {Carva}, \citenamefont {Vali\u{e}ka}, \citenamefont
    {Dole\u{z}al}, \citenamefont {Kratochv\'{i}lov\'{a}}, \citenamefont
    {Barnett}, \citenamefont {Proschek}, \citenamefont {Prokle\u{s}ka},
    \citenamefont {Kadlec}, \citenamefont {Ku\u{z}el}, \citenamefont {Colman},\
    and\ \citenamefont {Kamba}}]{Bastien2024}%
    \BibitemOpen
    \bibfield  {author} {\bibinfo {author} {\bibfnamefont {G.}~\bibnamefont
    {Bastien}}, \bibinfo {author} {\bibfnamefont {D.}~\bibnamefont
    {Rep{\v{c}}ek}}, \bibinfo {author} {\bibfnamefont {A.}~\bibnamefont
    {Eli{\'a}{\v{s}}}}, \bibinfo {author} {\bibfnamefont {A.}~\bibnamefont
    {Kancko}}, \bibinfo {author} {\bibfnamefont {Q.}~\bibnamefont {Courtade}},
    \bibinfo {author} {\bibfnamefont {T.}~\bibnamefont {Haidamak}}, \bibinfo
    {author} {\bibfnamefont {M.}~\bibnamefont {Savinov}}, \bibinfo {author}
    {\bibfnamefont {V.}~\bibnamefont {Bovtun}}, \bibinfo {author} {\bibfnamefont
    {M.}~\bibnamefont {Kempa}}, \bibinfo {author} {\bibfnamefont
    {K.}~\bibnamefont {Carva}}, \bibinfo {author} {\bibfnamefont
    {M.}~\bibnamefont {Vali\u{e}ka}}, \bibinfo {author} {\bibfnamefont
    {P.}~\bibnamefont {Dole\u{z}al}}, \bibinfo {author} {\bibfnamefont
    {M.}~\bibnamefont {Kratochv\'{i}lov\'{a}}}, \bibinfo {author} {\bibfnamefont
    {S.~A.}\ \bibnamefont {Barnett}}, \bibinfo {author} {\bibfnamefont
    {P.}~\bibnamefont {Proschek}}, \bibinfo {author} {\bibfnamefont
    {J.}~\bibnamefont {Prokle\u{s}ka}}, \bibinfo {author} {\bibfnamefont
    {C.}~\bibnamefont {Kadlec}}, \bibinfo {author} {\bibfnamefont
    {P.}~\bibnamefont {Ku\u{z}el}}, \bibinfo {author} {\bibfnamefont {R.~H.}\
    \bibnamefont {Colman}},\ and\ \bibinfo {author} {\bibfnamefont
    {S.}~\bibnamefont {Kamba}},\ }\bibfield  {title} {\bibinfo {title} {A
    frustrated antipolar phase analogous to classical spin liquids},\ }\href
    {https://doi.org/10.1002/adma.202410282} {\bibfield  {journal} {\bibinfo
    {journal} {Adv. Mater.}\ }\textbf {\bibinfo {volume} {36}},\ \bibinfo {pages}
    {2410282} (\bibinfo {year} {2024})}\BibitemShut {NoStop}%
  \bibitem [{\citenamefont {Zhu}\ \emph {et~al.}(2025)\citenamefont {Zhu},
    \citenamefont {Chinellato}, \citenamefont {Romerio}, \citenamefont {Murai},
    \citenamefont {Ohira-Kawamura}, \citenamefont {Balz}, \citenamefont {Yan},
    \citenamefont {Gvasaliya}, \citenamefont {Kato}, \citenamefont {Batista},\
    and\ \citenamefont {Zheludev}}]{Zhu2025}%
    \BibitemOpen
    \bibfield  {author} {\bibinfo {author} {\bibfnamefont {M.}~\bibnamefont
    {Zhu}}, \bibinfo {author} {\bibfnamefont {L.~M.}\ \bibnamefont {Chinellato}},
    \bibinfo {author} {\bibfnamefont {V.}~\bibnamefont {Romerio}}, \bibinfo
    {author} {\bibfnamefont {N.}~\bibnamefont {Murai}}, \bibinfo {author}
    {\bibfnamefont {S.}~\bibnamefont {Ohira-Kawamura}}, \bibinfo {author}
    {\bibfnamefont {C.}~\bibnamefont {Balz}}, \bibinfo {author} {\bibfnamefont
    {Z.}~\bibnamefont {Yan}}, \bibinfo {author} {\bibfnamefont {S.}~\bibnamefont
    {Gvasaliya}}, \bibinfo {author} {\bibfnamefont {Y.}~\bibnamefont {Kato}},
    \bibinfo {author} {\bibfnamefont {C.}~\bibnamefont {Batista}},\ and\ \bibinfo
    {author} {\bibfnamefont {A.}~\bibnamefont {Zheludev}},\ }\bibfield  {title}
    {\bibinfo {title} {Wannier states and spin supersolid physics in the
    triangular antiferromagnet {K$_2$Co(SeO$_3$)$_2$}},\ }\href
    {https://doi.org/10.1038/s41535-025-00791-2} {\bibfield  {journal} {\bibinfo
    {journal} {npj Quantum Mater.}\ }\textbf {\bibinfo {volume} {10}},\ \bibinfo
    {pages} {74} (\bibinfo {year} {2025})}\BibitemShut {NoStop}%
  \bibitem [{\citenamefont {Ottaviano}\ \emph {et~al.}(2003)\citenamefont
    {Ottaviano}, \citenamefont {Ressel}, \citenamefont {Di~Teodoro},
    \citenamefont {Profeta}, \citenamefont {Santucci}, \citenamefont {Ch\'ab},\
    and\ \citenamefont {Prince}}]{Ottaviano2003}%
    \BibitemOpen
    \bibfield  {author} {\bibinfo {author} {\bibfnamefont {L.}~\bibnamefont
    {Ottaviano}}, \bibinfo {author} {\bibfnamefont {B.}~\bibnamefont {Ressel}},
    \bibinfo {author} {\bibfnamefont {C.}~\bibnamefont {Di~Teodoro}}, \bibinfo
    {author} {\bibfnamefont {G.}~\bibnamefont {Profeta}}, \bibinfo {author}
    {\bibfnamefont {S.}~\bibnamefont {Santucci}}, \bibinfo {author}
    {\bibfnamefont {V.}~\bibnamefont {Ch\'ab}},\ and\ \bibinfo {author}
    {\bibfnamefont {K.~C.}\ \bibnamefont {Prince}},\ }\bibfield  {title}
    {\bibinfo {title} {Short-range order in two-dimensional binary alloys},\
    }\href {https://doi.org/10.1103/PhysRevB.67.045401} {\bibfield  {journal}
    {\bibinfo  {journal} {Phys. Rev. B}\ }\textbf {\bibinfo {volume} {67}},\
    \bibinfo {pages} {045401} (\bibinfo {year} {2003})}\BibitemShut {NoStop}%
  \bibitem [{\citenamefont {Azizi}\ \emph {et~al.}(2020)\citenamefont {Azizi},
    \citenamefont {Dogan}, \citenamefont {Cain}, \citenamefont {Eskandari},
    \citenamefont {Yu}, \citenamefont {Glazer}, \citenamefont {Cohen},\ and\
    \citenamefont {Zettl}}]{Azizi2020}%
    \BibitemOpen
    \bibfield  {author} {\bibinfo {author} {\bibfnamefont {A.}~\bibnamefont
    {Azizi}}, \bibinfo {author} {\bibfnamefont {M.}~\bibnamefont {Dogan}},
    \bibinfo {author} {\bibfnamefont {J.~D.}\ \bibnamefont {Cain}}, \bibinfo
    {author} {\bibfnamefont {R.}~\bibnamefont {Eskandari}}, \bibinfo {author}
    {\bibfnamefont {X.}~\bibnamefont {Yu}}, \bibinfo {author} {\bibfnamefont
    {E.~C.}\ \bibnamefont {Glazer}}, \bibinfo {author} {\bibfnamefont {M.~L.}\
    \bibnamefont {Cohen}},\ and\ \bibinfo {author} {\bibfnamefont
    {A.}~\bibnamefont {Zettl}},\ }\bibfield  {title} {\bibinfo {title}
    {Frustration and atomic ordering in a monolayer semiconductor alloy},\ }\href
    {https://doi.org/10.1103/PhysRevLett.124.096101} {\bibfield  {journal}
    {\bibinfo  {journal} {Phys. Rev. Lett.}\ }\textbf {\bibinfo {volume} {124}},\
    \bibinfo {pages} {096101} (\bibinfo {year} {2020})}\BibitemShut {NoStop}%
  \bibitem [{\citenamefont {Charra}\ and\ \citenamefont
    {Cousty}(1998)}]{Charra1998}%
    \BibitemOpen
    \bibfield  {author} {\bibinfo {author} {\bibfnamefont {F.}~\bibnamefont
    {Charra}}\ and\ \bibinfo {author} {\bibfnamefont {J.}~\bibnamefont
    {Cousty}},\ }\bibfield  {title} {\bibinfo {title} {Surface-induced chirality
    in a self-assembled monolayer of discotic liquid crystal},\ }\href
    {https://doi.org/10.1103/PhysRevLett.80.1682} {\bibfield  {journal} {\bibinfo
     {journal} {Phys. Rev. Lett.}\ }\textbf {\bibinfo {volume} {80}},\ \bibinfo
    {pages} {1682} (\bibinfo {year} {1998})}\BibitemShut {NoStop}%
  \bibitem [{\citenamefont {Alfonso-Moro}\ \emph {et~al.}(2023)\citenamefont
    {Alfonso-Moro}, \citenamefont {Guisset}, \citenamefont {David}, \citenamefont
    {Canals}, \citenamefont {Coraux},\ and\ \citenamefont
    {Rougemaille}}]{AlfonsoMoro2023}%
    \BibitemOpen
    \bibfield  {author} {\bibinfo {author} {\bibfnamefont {M.}~\bibnamefont
    {Alfonso-Moro}}, \bibinfo {author} {\bibfnamefont {V.}~\bibnamefont
    {Guisset}}, \bibinfo {author} {\bibfnamefont {P.}~\bibnamefont {David}},
    \bibinfo {author} {\bibfnamefont {B.}~\bibnamefont {Canals}}, \bibinfo
    {author} {\bibfnamefont {J.}~\bibnamefont {Coraux}},\ and\ \bibinfo {author}
    {\bibfnamefont {N.}~\bibnamefont {Rougemaille}},\ }\bibfield  {title}
    {\bibinfo {title} {Geometrical frustration, correlated disorder, and emerging
    order in a corrugated {C}$_{60}$ monolayer},\ }\href
    {https://doi.org/10.1103/PhysRevLett.131.186201} {\bibfield  {journal}
    {\bibinfo  {journal} {Phys. Rev. Lett.}\ }\textbf {\bibinfo {volume} {131}},\
    \bibinfo {pages} {186201} (\bibinfo {year} {2023})}\BibitemShut {NoStop}%
  \bibitem [{\citenamefont {Davidovi\'c}\ \emph {et~al.}(1997)\citenamefont
    {Davidovi\'c}, \citenamefont {Kumar}, \citenamefont {Reich}, \citenamefont
    {Siegel}, \citenamefont {Field}, \citenamefont {Tiberio}, \citenamefont
    {Hey},\ and\ \citenamefont {Ploog}}]{Davidovic1997}%
    \BibitemOpen
    \bibfield  {author} {\bibinfo {author} {\bibfnamefont {D.}~\bibnamefont
    {Davidovi\'c}}, \bibinfo {author} {\bibfnamefont {S.}~\bibnamefont {Kumar}},
    \bibinfo {author} {\bibfnamefont {D.~H.}\ \bibnamefont {Reich}}, \bibinfo
    {author} {\bibfnamefont {J.}~\bibnamefont {Siegel}}, \bibinfo {author}
    {\bibfnamefont {S.~B.}\ \bibnamefont {Field}}, \bibinfo {author}
    {\bibfnamefont {R.~C.}\ \bibnamefont {Tiberio}}, \bibinfo {author}
    {\bibfnamefont {R.}~\bibnamefont {Hey}},\ and\ \bibinfo {author}
    {\bibfnamefont {K.}~\bibnamefont {Ploog}},\ }\bibfield  {title} {\bibinfo
    {title} {Magnetic correlations, geometrical frustration, and tunable disorder
    in arrays of superconducting rings},\ }\href
    {https://doi.org/10.1103/PhysRevB.55.6518} {\bibfield  {journal} {\bibinfo
    {journal} {Phys. Rev. B}\ }\textbf {\bibinfo {volume} {55}},\ \bibinfo
    {pages} {6518} (\bibinfo {year} {1997})}\BibitemShut {NoStop}%
  \bibitem [{\citenamefont {Wang}\ \emph {et~al.}(2006)\citenamefont {Wang},
    \citenamefont {Nisoli}, \citenamefont {Freitas}, \citenamefont {Li},
    \citenamefont {McConville}, \citenamefont {Cooley}, \citenamefont {Lund},
    \citenamefont {Samarth}, \citenamefont {Leighton}, \citenamefont {Crespi},\
    and\ \citenamefont {Schiffer}}]{Wang2006}%
    \BibitemOpen
    \bibfield  {author} {\bibinfo {author} {\bibfnamefont {R.}~\bibnamefont
    {Wang}}, \bibinfo {author} {\bibfnamefont {C.}~\bibnamefont {Nisoli}},
    \bibinfo {author} {\bibfnamefont {R.~S.~d.}\ \bibnamefont {Freitas}},
    \bibinfo {author} {\bibfnamefont {J.}~\bibnamefont {Li}}, \bibinfo {author}
    {\bibfnamefont {W.}~\bibnamefont {McConville}}, \bibinfo {author}
    {\bibfnamefont {B.}~\bibnamefont {Cooley}}, \bibinfo {author} {\bibfnamefont
    {M.}~\bibnamefont {Lund}}, \bibinfo {author} {\bibfnamefont {N.}~\bibnamefont
    {Samarth}}, \bibinfo {author} {\bibfnamefont {C.}~\bibnamefont {Leighton}},
    \bibinfo {author} {\bibfnamefont {V.}~\bibnamefont {Crespi}},\ and\ \bibinfo
    {author} {\bibfnamefont {P.}~\bibnamefont {Schiffer}},\ }\bibfield  {title}
    {\bibinfo {title} {Artificial `spin ice' in a geometrically frustrated
    lattice of nanoscale ferromagnetic islands},\ }\href
    {https://doi.org/10.1038/nature04447} {\bibfield  {journal} {\bibinfo
    {journal} {Nature}\ }\textbf {\bibinfo {volume} {439}},\ \bibinfo {pages}
    {303} (\bibinfo {year} {2006})}\BibitemShut {NoStop}%
  \bibitem [{\citenamefont {Nisoli}\ \emph {et~al.}(2013)\citenamefont {Nisoli},
    \citenamefont {Moessner},\ and\ \citenamefont {Schiffer}}]{Nisoli2013}%
    \BibitemOpen
    \bibfield  {author} {\bibinfo {author} {\bibfnamefont {C.}~\bibnamefont
    {Nisoli}}, \bibinfo {author} {\bibfnamefont {R.}~\bibnamefont {Moessner}},\
    and\ \bibinfo {author} {\bibfnamefont {P.}~\bibnamefont {Schiffer}},\
    }\bibfield  {title} {\bibinfo {title} {Colloquium: Artificial spin ice:
    Designing and imaging magnetic frustration},\ }\href
    {https://doi.org/10.1103/RevModPhys.85.1473} {\bibfield  {journal} {\bibinfo
    {journal} {Rev. Mod. Phys.}\ }\textbf {\bibinfo {volume} {85}},\ \bibinfo
    {pages} {1473} (\bibinfo {year} {2013})}\BibitemShut {NoStop}%
  \bibitem [{\citenamefont {Ortiz-Ambriz}\ \emph {et~al.}(2019)\citenamefont
    {Ortiz-Ambriz}, \citenamefont {Nisoli}, \citenamefont {Reichhardt},
    \citenamefont {Reichhardt},\ and\ \citenamefont {Tierno}}]{Tierno2019}%
    \BibitemOpen
    \bibfield  {author} {\bibinfo {author} {\bibfnamefont {A.}~\bibnamefont
    {Ortiz-Ambriz}}, \bibinfo {author} {\bibfnamefont {C.}~\bibnamefont
    {Nisoli}}, \bibinfo {author} {\bibfnamefont {C.}~\bibnamefont {Reichhardt}},
    \bibinfo {author} {\bibfnamefont {C.~J.~O.}\ \bibnamefont {Reichhardt}},\
    and\ \bibinfo {author} {\bibfnamefont {P.}~\bibnamefont {Tierno}},\
    }\bibfield  {title} {\bibinfo {title} {Colloquium: Ice rule and emergent
    frustration in particle ice and beyond},\ }\href
    {https://doi.org/10.1103/RevModPhys.91.041003} {\bibfield  {journal}
    {\bibinfo  {journal} {Rev. Mod. Phys.}\ }\textbf {\bibinfo {volume} {91}},\
    \bibinfo {pages} {041003} (\bibinfo {year} {2019})}\BibitemShut {NoStop}%
  \bibitem [{\citenamefont {Rougemaille}\ and\ \citenamefont
    {Canals}(2019)}]{Rougemaille2019}%
    \BibitemOpen
    \bibfield  {author} {\bibinfo {author} {\bibfnamefont {N.}~\bibnamefont
    {Rougemaille}}\ and\ \bibinfo {author} {\bibfnamefont {B.}~\bibnamefont
    {Canals}},\ }\bibfield  {title} {\bibinfo {title} {Cooperative magnetic
    phenomena in artificial spin systems: spin liquids, coulomb phase and
    fragmentation of magnetism--a colloquium},\ }\href
    {https://doi.org/10.1140/epjb/e2018-90346-7} {\bibfield  {journal} {\bibinfo
    {journal} {Eur. Phys. J. B}\ }\textbf {\bibinfo {volume} {92}},\ \bibinfo
    {pages} {62} (\bibinfo {year} {2019})}\BibitemShut {NoStop}%
  \bibitem [{\citenamefont {Skj{\ae}rv{\o}}\ \emph {et~al.}(2020)\citenamefont
    {Skj{\ae}rv{\o}}, \citenamefont {Marrows}, \citenamefont {Stamps},\ and\
    \citenamefont {Heyderman}}]{Skjaervo2020}%
    \BibitemOpen
    \bibfield  {author} {\bibinfo {author} {\bibfnamefont {S.~H.}\ \bibnamefont
    {Skj{\ae}rv{\o}}}, \bibinfo {author} {\bibfnamefont {C.~H.}\ \bibnamefont
    {Marrows}}, \bibinfo {author} {\bibfnamefont {R.~L.}\ \bibnamefont
    {Stamps}},\ and\ \bibinfo {author} {\bibfnamefont {L.~J.}\ \bibnamefont
    {Heyderman}},\ }\bibfield  {title} {\bibinfo {title} {Advances in artificial
    spin ice},\ }\href {https://doi.org/10.1038/s42254-019-0118-3} {\bibfield
    {journal} {\bibinfo  {journal} {Nat. Rev. Phys.}\ }\textbf {\bibinfo {volume}
    {2}},\ \bibinfo {pages} {13} (\bibinfo {year} {2020})}\BibitemShut {NoStop}%
  \bibitem [{\citenamefont {Tanaka}\ \emph {et~al.}(2005)\citenamefont {Tanaka},
    \citenamefont {Saitoh}, \citenamefont {Miyajima}, \citenamefont {Yamaoka},\
    and\ \citenamefont {Iye}}]{Tanaka2005}%
    \BibitemOpen
    \bibfield  {author} {\bibinfo {author} {\bibfnamefont {M.}~\bibnamefont
    {Tanaka}}, \bibinfo {author} {\bibfnamefont {E.}~\bibnamefont {Saitoh}},
    \bibinfo {author} {\bibfnamefont {H.}~\bibnamefont {Miyajima}}, \bibinfo
    {author} {\bibfnamefont {T.}~\bibnamefont {Yamaoka}},\ and\ \bibinfo {author}
    {\bibfnamefont {Y.}~\bibnamefont {Iye}},\ }\bibfield  {title} {\bibinfo
    {title} {Domain structures and magnetic ice-order in {NiFe} nano-network with
    honeycomb structure},\ }\href {https://doi.org/10.1063/1.1854572} {\bibfield
    {journal} {\bibinfo  {journal} {J. Appl. Phys.}\ }\textbf {\bibinfo {volume}
    {97}},\ \bibinfo {pages} {10J710} (\bibinfo {year} {2005})}\BibitemShut
    {NoStop}%
  \bibitem [{\citenamefont {Qi}\ \emph {et~al.}(2008)\citenamefont {Qi},
    \citenamefont {Brintlinger},\ and\ \citenamefont {Cumings}}]{Qi2008}%
    \BibitemOpen
    \bibfield  {author} {\bibinfo {author} {\bibfnamefont {Y.}~\bibnamefont
    {Qi}}, \bibinfo {author} {\bibfnamefont {T.}~\bibnamefont {Brintlinger}},\
    and\ \bibinfo {author} {\bibfnamefont {J.}~\bibnamefont {Cumings}},\
    }\bibfield  {title} {\bibinfo {title} {Direct observation of the ice rule in
    an artificial kagome spin ice},\ }\href
    {https://doi.org/10.1103/PhysRevB.77.094418} {\bibfield  {journal} {\bibinfo
    {journal} {Phys. Rev. B}\ }\textbf {\bibinfo {volume} {77}},\ \bibinfo
    {pages} {094418} (\bibinfo {year} {2008})}\BibitemShut {NoStop}%
  \bibitem [{\citenamefont {Rougemaille}\ \emph {et~al.}(2011)\citenamefont
    {Rougemaille}, \citenamefont {Montaigne}, \citenamefont {Canals},
    \citenamefont {Duluard}, \citenamefont {Lacour}, \citenamefont {Hehn},
    \citenamefont {Belkhou}, \citenamefont {Fruchart}, \citenamefont
    {El~Moussaoui}, \citenamefont {Bendounan},\ and\ \citenamefont
    {Maccherozzi}}]{Rougemaille2011}%
    \BibitemOpen
    \bibfield  {author} {\bibinfo {author} {\bibfnamefont {N.}~\bibnamefont
    {Rougemaille}}, \bibinfo {author} {\bibfnamefont {F.}~\bibnamefont
    {Montaigne}}, \bibinfo {author} {\bibfnamefont {B.}~\bibnamefont {Canals}},
    \bibinfo {author} {\bibfnamefont {A.}~\bibnamefont {Duluard}}, \bibinfo
    {author} {\bibfnamefont {D.}~\bibnamefont {Lacour}}, \bibinfo {author}
    {\bibfnamefont {M.}~\bibnamefont {Hehn}}, \bibinfo {author} {\bibfnamefont
    {R.}~\bibnamefont {Belkhou}}, \bibinfo {author} {\bibfnamefont
    {O.}~\bibnamefont {Fruchart}}, \bibinfo {author} {\bibfnamefont
    {S.}~\bibnamefont {El~Moussaoui}}, \bibinfo {author} {\bibfnamefont
    {A.}~\bibnamefont {Bendounan}},\ and\ \bibinfo {author} {\bibfnamefont
    {F.}~\bibnamefont {Maccherozzi}},\ }\bibfield  {title} {\bibinfo {title}
    {Artificial kagome arrays of nanomagnets: A frozen dipolar spin ice},\ }\href
    {https://doi.org/10.1103/PhysRevLett.106.057209} {\bibfield  {journal}
    {\bibinfo  {journal} {Phys. Rev. Lett.}\ }\textbf {\bibinfo {volume} {106}},\
    \bibinfo {pages} {057209} (\bibinfo {year} {2011})}\BibitemShut {NoStop}%
  \bibitem [{\citenamefont {Zhang}\ \emph {et~al.}(2013)\citenamefont {Zhang},
    \citenamefont {Gilbert}, \citenamefont {Nisoli}, \citenamefont {Chern},
    \citenamefont {Erickson}, \citenamefont {O’brien}, \citenamefont
    {Leighton}, \citenamefont {Lammert}, \citenamefont {Crespi},\ and\
    \citenamefont {Schiffer}}]{Zhang2013}%
    \BibitemOpen
    \bibfield  {author} {\bibinfo {author} {\bibfnamefont {S.}~\bibnamefont
    {Zhang}}, \bibinfo {author} {\bibfnamefont {I.}~\bibnamefont {Gilbert}},
    \bibinfo {author} {\bibfnamefont {C.}~\bibnamefont {Nisoli}}, \bibinfo
    {author} {\bibfnamefont {G.-W.}\ \bibnamefont {Chern}}, \bibinfo {author}
    {\bibfnamefont {M.~J.}\ \bibnamefont {Erickson}}, \bibinfo {author}
    {\bibfnamefont {L.}~\bibnamefont {O’brien}}, \bibinfo {author}
    {\bibfnamefont {C.}~\bibnamefont {Leighton}}, \bibinfo {author}
    {\bibfnamefont {P.~E.}\ \bibnamefont {Lammert}}, \bibinfo {author}
    {\bibfnamefont {V.~H.}\ \bibnamefont {Crespi}},\ and\ \bibinfo {author}
    {\bibfnamefont {P.}~\bibnamefont {Schiffer}},\ }\bibfield  {title} {\bibinfo
    {title} {Crystallites of magnetic charges in artificial spin ice},\ }\href
    {https://doi.org/10.1038/nature12399} {\bibfield  {journal} {\bibinfo
    {journal} {Nature}\ }\textbf {\bibinfo {volume} {500}},\ \bibinfo {pages}
    {553} (\bibinfo {year} {2013})}\BibitemShut {NoStop}%
  \bibitem [{\citenamefont {Anghinolfi}\ \emph {et~al.}(2015)\citenamefont
    {Anghinolfi}, \citenamefont {Luetkens}, \citenamefont {Perron}, \citenamefont
    {Flokstra}, \citenamefont {Sendetskyi}, \citenamefont {Suter}, \citenamefont
    {Prokscha}, \citenamefont {Derlet}, \citenamefont {Lee},\ and\ \citenamefont
    {Heyderman}}]{Anghinolfi2015}%
    \BibitemOpen
    \bibfield  {author} {\bibinfo {author} {\bibfnamefont {L.}~\bibnamefont
    {Anghinolfi}}, \bibinfo {author} {\bibfnamefont {H.}~\bibnamefont
    {Luetkens}}, \bibinfo {author} {\bibfnamefont {J.}~\bibnamefont {Perron}},
    \bibinfo {author} {\bibfnamefont {M.}~\bibnamefont {Flokstra}}, \bibinfo
    {author} {\bibfnamefont {O.}~\bibnamefont {Sendetskyi}}, \bibinfo {author}
    {\bibfnamefont {A.}~\bibnamefont {Suter}}, \bibinfo {author} {\bibfnamefont
    {T.}~\bibnamefont {Prokscha}}, \bibinfo {author} {\bibfnamefont
    {P.}~\bibnamefont {Derlet}}, \bibinfo {author} {\bibfnamefont
    {S.}~\bibnamefont {Lee}},\ and\ \bibinfo {author} {\bibfnamefont
    {L.}~\bibnamefont {Heyderman}},\ }\bibfield  {title} {\bibinfo {title}
    {Thermodynamic phase transitions in a frustrated magnetic metamaterial},\
    }\href {https://doi.org/10.1038/ncomms9278} {\bibfield  {journal} {\bibinfo
    {journal} {Nat. Commun.}\ }\textbf {\bibinfo {volume} {6}},\ \bibinfo {pages}
    {8278} (\bibinfo {year} {2015})}\BibitemShut {NoStop}%
  \bibitem [{\citenamefont {Perrin}\ \emph {et~al.}(2016)\citenamefont {Perrin},
    \citenamefont {Canals},\ and\ \citenamefont {Rougemaille}}]{Perrin2016}%
    \BibitemOpen
    \bibfield  {author} {\bibinfo {author} {\bibfnamefont {Y.}~\bibnamefont
    {Perrin}}, \bibinfo {author} {\bibfnamefont {B.}~\bibnamefont {Canals}},\
    and\ \bibinfo {author} {\bibfnamefont {N.}~\bibnamefont {Rougemaille}},\
    }\bibfield  {title} {\bibinfo {title} {Extensive degeneracy, {Coulomb} phase
    and magnetic monopoles in artificial square ice},\ }\href
    {https://doi.org/10.1038/nature20155} {\bibfield  {journal} {\bibinfo
    {journal} {Nature}\ }\textbf {\bibinfo {volume} {540}},\ \bibinfo {pages}
    {410} (\bibinfo {year} {2016})}\BibitemShut {NoStop}%
  \bibitem [{\citenamefont {{\"O}stman}\ \emph
    {et~al.}(2018{\natexlab{a}})\citenamefont {{\"O}stman}, \citenamefont
    {Stopfel}, \citenamefont {Chioar}, \citenamefont {Arnalds}, \citenamefont
    {Stein}, \citenamefont {Kapaklis},\ and\ \citenamefont
    {Hj{\"o}rvarsson}}]{Ostman2018}%
    \BibitemOpen
    \bibfield  {author} {\bibinfo {author} {\bibfnamefont {E.}~\bibnamefont
    {{\"O}stman}}, \bibinfo {author} {\bibfnamefont {H.}~\bibnamefont {Stopfel}},
    \bibinfo {author} {\bibfnamefont {I.-A.}\ \bibnamefont {Chioar}}, \bibinfo
    {author} {\bibfnamefont {U.~B.}\ \bibnamefont {Arnalds}}, \bibinfo {author}
    {\bibfnamefont {A.}~\bibnamefont {Stein}}, \bibinfo {author} {\bibfnamefont
    {V.}~\bibnamefont {Kapaklis}},\ and\ \bibinfo {author} {\bibfnamefont
    {B.}~\bibnamefont {Hj{\"o}rvarsson}},\ }\bibfield  {title} {\bibinfo {title}
    {Interaction modifiers in artificial spin ices},\ }\href
    {https://doi.org/10.1038/s41567-017-0027-2} {\bibfield  {journal} {\bibinfo
    {journal} {Nat. Phys.}\ }\textbf {\bibinfo {volume} {14}},\ \bibinfo {pages}
    {375} (\bibinfo {year} {2018}{\natexlab{a}})}\BibitemShut {NoStop}%
  \bibitem [{\citenamefont {Sch\'anilec}\ \emph {et~al.}(2022)\citenamefont
    {Sch\'anilec}, \citenamefont {Brunn}, \citenamefont
    {Hor\'a\ifmmode~\check{c}\else \v{c}\fi{}ek}, \citenamefont {Kr\'atk\'y},
    \citenamefont {Meluz\'{\i}n}, \citenamefont {\ifmmode~\check{S}\else
    \v{S}\fi{}ikola}, \citenamefont {Canals},\ and\ \citenamefont
    {Rougemaille}}]{Schanilec2022}%
    \BibitemOpen
    \bibfield  {author} {\bibinfo {author} {\bibfnamefont {V.}~\bibnamefont
    {Sch\'anilec}}, \bibinfo {author} {\bibfnamefont {O.}~\bibnamefont {Brunn}},
    \bibinfo {author} {\bibfnamefont {M.}~\bibnamefont
    {Hor\'a\ifmmode~\check{c}\else \v{c}\fi{}ek}}, \bibinfo {author}
    {\bibfnamefont {S.}~\bibnamefont {Kr\'atk\'y}}, \bibinfo {author}
    {\bibfnamefont {P.}~\bibnamefont {Meluz\'{\i}n}}, \bibinfo {author}
    {\bibfnamefont {T.}~\bibnamefont {\ifmmode~\check{S}\else \v{S}\fi{}ikola}},
    \bibinfo {author} {\bibfnamefont {B.}~\bibnamefont {Canals}},\ and\ \bibinfo
    {author} {\bibfnamefont {N.}~\bibnamefont {Rougemaille}},\ }\bibfield
    {title} {\bibinfo {title} {Approaching the topological low-energy physics of
    the {$F$} model in a two-dimensional magnetic lattice},\ }\href
    {https://doi.org/10.1103/PhysRevLett.129.027202} {\bibfield  {journal}
    {\bibinfo  {journal} {Phys. Rev. Lett.}\ }\textbf {\bibinfo {volume} {129}},\
    \bibinfo {pages} {027202} (\bibinfo {year} {2022})}\BibitemShut {NoStop}%
  \bibitem [{\citenamefont {Arnalds}\ \emph {et~al.}(2016)\citenamefont
    {Arnalds}, \citenamefont {Chico}, \citenamefont {Stopfel}, \citenamefont
    {Kapaklis}, \citenamefont {B{\"a}renbold}, \citenamefont {Verschuuren},
    \citenamefont {Wolff}, \citenamefont {Neu}, \citenamefont {Bergman},\ and\
    \citenamefont {Hj{\"o}rvarsson}}]{Arnalds2016}%
    \BibitemOpen
    \bibfield  {author} {\bibinfo {author} {\bibfnamefont {U.~B.}\ \bibnamefont
    {Arnalds}}, \bibinfo {author} {\bibfnamefont {J.}~\bibnamefont {Chico}},
    \bibinfo {author} {\bibfnamefont {H.}~\bibnamefont {Stopfel}}, \bibinfo
    {author} {\bibfnamefont {V.}~\bibnamefont {Kapaklis}}, \bibinfo {author}
    {\bibfnamefont {O.}~\bibnamefont {B{\"a}renbold}}, \bibinfo {author}
    {\bibfnamefont {M.~A.}\ \bibnamefont {Verschuuren}}, \bibinfo {author}
    {\bibfnamefont {U.}~\bibnamefont {Wolff}}, \bibinfo {author} {\bibfnamefont
    {V.}~\bibnamefont {Neu}}, \bibinfo {author} {\bibfnamefont {A.}~\bibnamefont
    {Bergman}},\ and\ \bibinfo {author} {\bibfnamefont {B.}~\bibnamefont
    {Hj{\"o}rvarsson}},\ }\bibfield  {title} {\bibinfo {title} {A new look on the
    two-dimensional {Ising} model: thermal artificial spins},\ }\href
    {https://doi.org/10.1088/1367-2630/18/2/023008} {\bibfield  {journal}
    {\bibinfo  {journal} {New J. Phys.}\ }\textbf {\bibinfo {volume} {18}},\
    \bibinfo {pages} {023008} (\bibinfo {year} {2016})}\BibitemShut {NoStop}%
  \bibitem [{\citenamefont {Nguyen}\ \emph {et~al.}(2017)\citenamefont {Nguyen},
    \citenamefont {Perrin}, \citenamefont {Le~Denmat}, \citenamefont {Canals},\
    and\ \citenamefont {Rougemaille}}]{Nguyen2017}%
    \BibitemOpen
    \bibfield  {author} {\bibinfo {author} {\bibfnamefont {V.-D.}\ \bibnamefont
    {Nguyen}}, \bibinfo {author} {\bibfnamefont {Y.}~\bibnamefont {Perrin}},
    \bibinfo {author} {\bibfnamefont {S.}~\bibnamefont {Le~Denmat}}, \bibinfo
    {author} {\bibfnamefont {B.}~\bibnamefont {Canals}},\ and\ \bibinfo {author}
    {\bibfnamefont {N.}~\bibnamefont {Rougemaille}},\ }\bibfield  {title}
    {\bibinfo {title} {Competing interactions in artificial spin chains},\ }\href
    {https://doi.org/10.1103/PhysRevB.96.014402} {\bibfield  {journal} {\bibinfo
    {journal} {Phys. Rev. B}\ }\textbf {\bibinfo {volume} {96}},\ \bibinfo
    {pages} {014402} (\bibinfo {year} {2017})}\BibitemShut {NoStop}%
  \bibitem [{\citenamefont {{\"O}stman}\ \emph
    {et~al.}(2018{\natexlab{b}})\citenamefont {{\"O}stman}, \citenamefont
    {Arnalds}, \citenamefont {Kapaklis}, \citenamefont {Taroni},\ and\
    \citenamefont {Hj{\"o}rvarsson}}]{Ostman2018b}%
    \BibitemOpen
    \bibfield  {author} {\bibinfo {author} {\bibfnamefont {E.}~\bibnamefont
    {{\"O}stman}}, \bibinfo {author} {\bibfnamefont {U.~B.}\ \bibnamefont
    {Arnalds}}, \bibinfo {author} {\bibfnamefont {V.}~\bibnamefont {Kapaklis}},
    \bibinfo {author} {\bibfnamefont {A.}~\bibnamefont {Taroni}},\ and\ \bibinfo
    {author} {\bibfnamefont {B.}~\bibnamefont {Hj{\"o}rvarsson}},\ }\bibfield
    {title} {\bibinfo {title} {Ising-like behaviour of mesoscopic magnetic
    chains},\ }\href {https://doi.org/10.1088/1361-648X/aad0c1} {\bibfield
    {journal} {\bibinfo  {journal} {J. Phys.: Condens. Matter}\ }\textbf
    {\bibinfo {volume} {30}},\ \bibinfo {pages} {365301} (\bibinfo {year}
    {2018}{\natexlab{b}})}\BibitemShut {NoStop}%
  \bibitem [{\citenamefont {Farhan}\ \emph {et~al.}(2019)\citenamefont {Farhan},
    \citenamefont {Saccone}, \citenamefont {Petersen}, \citenamefont {Dhuey},
    \citenamefont {Chopdekar}, \citenamefont {Huang}, \citenamefont {Kent},
    \citenamefont {Chen}, \citenamefont {Alava}, \citenamefont {Lippert},
    \citenamefont {Scholl},\ and\ \citenamefont {van Dijken}}]{Farhan2019}%
    \BibitemOpen
    \bibfield  {author} {\bibinfo {author} {\bibfnamefont {A.}~\bibnamefont
    {Farhan}}, \bibinfo {author} {\bibfnamefont {M.}~\bibnamefont {Saccone}},
    \bibinfo {author} {\bibfnamefont {C.~F.}\ \bibnamefont {Petersen}}, \bibinfo
    {author} {\bibfnamefont {S.}~\bibnamefont {Dhuey}}, \bibinfo {author}
    {\bibfnamefont {R.~V.}\ \bibnamefont {Chopdekar}}, \bibinfo {author}
    {\bibfnamefont {Y.-L.}\ \bibnamefont {Huang}}, \bibinfo {author}
    {\bibfnamefont {N.}~\bibnamefont {Kent}}, \bibinfo {author} {\bibfnamefont
    {Z.}~\bibnamefont {Chen}}, \bibinfo {author} {\bibfnamefont {M.~J.}\
    \bibnamefont {Alava}}, \bibinfo {author} {\bibfnamefont {T.}~\bibnamefont
    {Lippert}}, \bibinfo {author} {\bibfnamefont {A.}~\bibnamefont {Scholl}},\
    and\ \bibinfo {author} {\bibfnamefont {S.}~\bibnamefont {van Dijken}},\
    }\bibfield  {title} {\bibinfo {title} {Emergent magnetic monopole dynamics in
    macroscopically degenerate artificial spin ice},\ }\href
    {https://doi.org/10.1126/sciadv.aav6380} {\bibfield  {journal} {\bibinfo
    {journal} {Sci. Adv.}\ }\textbf {\bibinfo {volume} {5}},\ \bibinfo {pages}
    {eaav6380} (\bibinfo {year} {2019})}\BibitemShut {NoStop}%
  \bibitem [{\citenamefont {May}\ \emph {et~al.}(2021)\citenamefont {May},
    \citenamefont {Saccone}, \citenamefont {van~den Berg}, \citenamefont {Askey},
    \citenamefont {Hunt},\ and\ \citenamefont {Ladak}}]{May2021}%
    \BibitemOpen
    \bibfield  {author} {\bibinfo {author} {\bibfnamefont {A.}~\bibnamefont
    {May}}, \bibinfo {author} {\bibfnamefont {M.}~\bibnamefont {Saccone}},
    \bibinfo {author} {\bibfnamefont {A.}~\bibnamefont {van~den Berg}}, \bibinfo
    {author} {\bibfnamefont {J.}~\bibnamefont {Askey}}, \bibinfo {author}
    {\bibfnamefont {M.}~\bibnamefont {Hunt}},\ and\ \bibinfo {author}
    {\bibfnamefont {S.}~\bibnamefont {Ladak}},\ }\bibfield  {title} {\bibinfo
    {title} {Magnetic charge propagation upon a {3D} artificial spin-ice},\
    }\href {https://doi.org/10.1038/s41467-021-23480-7} {\bibfield  {journal}
    {\bibinfo  {journal} {Nat. Commun.}\ }\textbf {\bibinfo {volume} {12}},\
    \bibinfo {pages} {3217} (\bibinfo {year} {2021})}\BibitemShut {NoStop}%
  \bibitem [{\citenamefont {Han}\ \emph {et~al.}(2008)\citenamefont {Han},
    \citenamefont {Shokef}, \citenamefont {Alsayed}, \citenamefont {Yunker},
    \citenamefont {Lubensky},\ and\ \citenamefont {Yodh}}]{Han2008}%
    \BibitemOpen
    \bibfield  {author} {\bibinfo {author} {\bibfnamefont {Y.}~\bibnamefont
    {Han}}, \bibinfo {author} {\bibfnamefont {Y.}~\bibnamefont {Shokef}},
    \bibinfo {author} {\bibfnamefont {A.~M.}\ \bibnamefont {Alsayed}}, \bibinfo
    {author} {\bibfnamefont {P.}~\bibnamefont {Yunker}}, \bibinfo {author}
    {\bibfnamefont {T.~C.}\ \bibnamefont {Lubensky}},\ and\ \bibinfo {author}
    {\bibfnamefont {A.~G.}\ \bibnamefont {Yodh}},\ }\bibfield  {title} {\bibinfo
    {title} {Geometric frustration in buckled colloidal monolayers},\ }\href
    {https://doi.org/10.1038/nature07595} {\bibfield  {journal} {\bibinfo
    {journal} {Nature}\ }\textbf {\bibinfo {volume} {456}},\ \bibinfo {pages}
    {898} (\bibinfo {year} {2008})}\BibitemShut {NoStop}%
  \bibitem [{\citenamefont {Farhan}\ \emph {et~al.}(2020)\citenamefont {Farhan},
    \citenamefont {Saccone}, \citenamefont {Petersen}, \citenamefont {Dhuey},
    \citenamefont {Hofhuis}, \citenamefont {Mansell}, \citenamefont {Chopdekar},
    \citenamefont {Scholl}, \citenamefont {Lippert},\ and\ \citenamefont
    {Van~Dijken}}]{Farhan2020}%
    \BibitemOpen
    \bibfield  {author} {\bibinfo {author} {\bibfnamefont {A.}~\bibnamefont
    {Farhan}}, \bibinfo {author} {\bibfnamefont {M.}~\bibnamefont {Saccone}},
    \bibinfo {author} {\bibfnamefont {C.~F.}\ \bibnamefont {Petersen}}, \bibinfo
    {author} {\bibfnamefont {S.}~\bibnamefont {Dhuey}}, \bibinfo {author}
    {\bibfnamefont {K.}~\bibnamefont {Hofhuis}}, \bibinfo {author} {\bibfnamefont
    {R.}~\bibnamefont {Mansell}}, \bibinfo {author} {\bibfnamefont {R.~V.}\
    \bibnamefont {Chopdekar}}, \bibinfo {author} {\bibfnamefont {A.}~\bibnamefont
    {Scholl}}, \bibinfo {author} {\bibfnamefont {T.}~\bibnamefont {Lippert}},\
    and\ \bibinfo {author} {\bibfnamefont {S.}~\bibnamefont {Van~Dijken}},\
    }\bibfield  {title} {\bibinfo {title} {Geometrical frustration and planar
    triangular antiferromagnetism in quasi-three-dimensional artificial spin
    architecture},\ }\href {https://doi.org/10.1103/PhysRevLett.125.267203}
    {\bibfield  {journal} {\bibinfo  {journal} {Phys. Rev. Lett.}\ }\textbf
    {\bibinfo {volume} {125}},\ \bibinfo {pages} {267203} (\bibinfo {year}
    {2020})}\BibitemShut {NoStop}%
  \bibitem [{\citenamefont {Pip}\ \emph {et~al.}(2021)\citenamefont {Pip},
    \citenamefont {Glavic}, \citenamefont {Skj{\ae}rv{\o}}, \citenamefont
    {Weber}, \citenamefont {Smerald}, \citenamefont {Zhernenkov}, \citenamefont
    {Leo}, \citenamefont {Mila}, \citenamefont {Philippe},\ and\ \citenamefont
    {Heyderman}}]{Pip2021}%
    \BibitemOpen
    \bibfield  {author} {\bibinfo {author} {\bibfnamefont {P.}~\bibnamefont
    {Pip}}, \bibinfo {author} {\bibfnamefont {A.}~\bibnamefont {Glavic}},
    \bibinfo {author} {\bibfnamefont {S.~H.}\ \bibnamefont {Skj{\ae}rv{\o}}},
    \bibinfo {author} {\bibfnamefont {A.}~\bibnamefont {Weber}}, \bibinfo
    {author} {\bibfnamefont {A.}~\bibnamefont {Smerald}}, \bibinfo {author}
    {\bibfnamefont {K.}~\bibnamefont {Zhernenkov}}, \bibinfo {author}
    {\bibfnamefont {N.}~\bibnamefont {Leo}}, \bibinfo {author} {\bibfnamefont
    {F.}~\bibnamefont {Mila}}, \bibinfo {author} {\bibfnamefont {L.}~\bibnamefont
    {Philippe}},\ and\ \bibinfo {author} {\bibfnamefont {L.~J.}\ \bibnamefont
    {Heyderman}},\ }\bibfield  {title} {\bibinfo {title} {Direct observation of
    spin correlations in an artificial triangular lattice {Ising} spin system
    with grazing-incidence small-angle neutron scattering},\ }\href
    {https://doi.org/10.1039/D1NH00043H} {\bibfield  {journal} {\bibinfo
    {journal} {Nanoscale Horiz.}\ }\textbf {\bibinfo {volume} {6}},\ \bibinfo
    {pages} {474} (\bibinfo {year} {2021})}\BibitemShut {NoStop}%
  \bibitem [{\citenamefont {Pac}\ \emph {et~al.}(2026)\citenamefont {Pac},
    \citenamefont {Macauley}, \citenamefont {Massey}, \citenamefont {Kurenkov},
    \citenamefont {Mila}, \citenamefont {Derlet},\ and\ \citenamefont
    {Heyderman}}]{Pac2025}%
    \BibitemOpen
    \bibfield  {author} {\bibinfo {author} {\bibfnamefont {A.}~\bibnamefont
    {Pac}}, \bibinfo {author} {\bibfnamefont {G.~M.}\ \bibnamefont {Macauley}},
    \bibinfo {author} {\bibfnamefont {J.~R.}\ \bibnamefont {Massey}}, \bibinfo
    {author} {\bibfnamefont {A.}~\bibnamefont {Kurenkov}}, \bibinfo {author}
    {\bibfnamefont {F.}~\bibnamefont {Mila}}, \bibinfo {author} {\bibfnamefont
    {P.~M.}\ \bibnamefont {Derlet}},\ and\ \bibinfo {author} {\bibfnamefont
    {L.~J.}\ \bibnamefont {Heyderman}},\ }\bibfield  {title} {\bibinfo {title}
    {Magnetic ordering in out-of-plane artificial spin systems based on the
    {Archimedean} lattices},\ }\href {https://doi.org/10.1103/n2ns-j2hp}
    {\bibfield  {journal} {\bibinfo  {journal} {Phys. Rev. B}\ }\textbf {\bibinfo
    {volume} {113}},\ \bibinfo {pages} {144403} (\bibinfo {year}
    {2026})}\BibitemShut {NoStop}%
  \bibitem [{\citenamefont {Wang}\ \emph {et~al.}(2025)\citenamefont {Wang},
    \citenamefont {Liu}, \citenamefont {Tu}, \citenamefont {Zhang}, \citenamefont
    {Gladilin},\ and\ \citenamefont {Ge}}]{Wang2025}%
    \BibitemOpen
    \bibfield  {author} {\bibinfo {author} {\bibfnamefont {K.}~\bibnamefont
    {Wang}}, \bibinfo {author} {\bibfnamefont {X.-J.}\ \bibnamefont {Liu}},
    \bibinfo {author} {\bibfnamefont {L.-M.}\ \bibnamefont {Tu}}, \bibinfo
    {author} {\bibfnamefont {J.-J.}\ \bibnamefont {Zhang}}, \bibinfo {author}
    {\bibfnamefont {V.~N.}\ \bibnamefont {Gladilin}},\ and\ \bibinfo {author}
    {\bibfnamefont {J.-Y.}\ \bibnamefont {Ge}},\ }\bibfield  {title} {\bibinfo
    {title} {Experimental realization of antiferromagnetic ising ground state on
    the triangular lattice},\ }\href {https://doi.org/10.1103/syny-gvt3}
    {\bibfield  {journal} {\bibinfo  {journal} {Phys. Rev. B}\ }\textbf {\bibinfo
    {volume} {111}},\ \bibinfo {pages} {224418} (\bibinfo {year}
    {2025})}\BibitemShut {NoStop}%
  \bibitem [{\citenamefont {Wannier}(1973)}]{Wannier1973}%
    \BibitemOpen
    \bibfield  {author} {\bibinfo {author} {\bibfnamefont {G.}~\bibnamefont
    {Wannier}},\ }\bibfield  {title} {\bibinfo {title} {Antiferromagnetism. the
    triangular ising net},\ }\href {https://doi.org/10.1103/physrevb.7.5017}
    {\bibfield  {journal} {\bibinfo  {journal} {Phys. Rev. B}\ }\textbf {\bibinfo
    {volume} {7}},\ \bibinfo {pages} {5017} (\bibinfo {year} {1973})}\BibitemShut
    {NoStop}%
  \bibitem [{\citenamefont {Ge}\ \emph {et~al.}(2025)\citenamefont {Ge},
    \citenamefont {Zhang}, \citenamefont {Tu},\ and\ \citenamefont
    {Gladilin}}]{Ge2025}%
    \BibitemOpen
    \bibfield  {author} {\bibinfo {author} {\bibfnamefont {J.-Y.}\ \bibnamefont
    {Ge}}, \bibinfo {author} {\bibfnamefont {J.-J.}\ \bibnamefont {Zhang}},
    \bibinfo {author} {\bibfnamefont {L.-M.}\ \bibnamefont {Tu}},\ and\ \bibinfo
    {author} {\bibfnamefont {V.~N.}\ \bibnamefont {Gladilin}},\ }\bibfield
    {title} {\bibinfo {title} {Growing crystalline artificial kagome ice at the
    macroscale},\ }\href
    {https://doi.org/https://doi.org/10.1016/j.newton.2025.100143} {\bibfield
    {journal} {\bibinfo  {journal} {Newton}\ }\textbf {\bibinfo {volume} {1}},\
    \bibinfo {pages} {100143} (\bibinfo {year} {2025})}\BibitemShut {NoStop}%
  \bibitem [{\citenamefont {Ortiz-Ambriz}\ and\ \citenamefont
    {Tierno}(2016)}]{Tierno2016}%
    \BibitemOpen
    \bibfield  {author} {\bibinfo {author} {\bibfnamefont {A.}~\bibnamefont
    {Ortiz-Ambriz}}\ and\ \bibinfo {author} {\bibfnamefont {P.}~\bibnamefont
    {Tierno}},\ }\bibfield  {title} {\bibinfo {title} {Engineering of frustration
    in colloidal artificial ices realized on microfeatured grooved lattices},\
    }\href {https://doi.org/10.1038/ncomms10575} {\bibfield  {journal} {\bibinfo
    {journal} {Nat. Commun.}\ }\textbf {\bibinfo {volume} {7}},\ \bibinfo {pages}
    {10575} (\bibinfo {year} {2016})}\BibitemShut {NoStop}%
  \bibitem [{\citenamefont {Olive}\ and\ \citenamefont
    {Molho}(1998)}]{Olive1998}%
    \BibitemOpen
    \bibfield  {author} {\bibinfo {author} {\bibfnamefont {E.}~\bibnamefont
    {Olive}}\ and\ \bibinfo {author} {\bibfnamefont {P.}~\bibnamefont {Molho}},\
    }\bibfield  {title} {\bibinfo {title} {Thermodynamic study of a lattice of
    compass needles in dipolar interaction},\ }\href
    {https://doi.org/10.1103/PhysRevB.58.9238} {\bibfield  {journal} {\bibinfo
    {journal} {Phys. Rev. B}\ }\textbf {\bibinfo {volume} {58}},\ \bibinfo
    {pages} {9238} (\bibinfo {year} {1998})}\BibitemShut {NoStop}%
  \bibitem [{\citenamefont {Vedmedenko}\ \emph {et~al.}(2003)\citenamefont
    {Vedmedenko}, \citenamefont {Oepen},\ and\ \citenamefont
    {Kirschner}}]{Kirschner2003}%
    \BibitemOpen
    \bibfield  {author} {\bibinfo {author} {\bibfnamefont {E.~Y.}\ \bibnamefont
    {Vedmedenko}}, \bibinfo {author} {\bibfnamefont {H.~P.}\ \bibnamefont
    {Oepen}},\ and\ \bibinfo {author} {\bibfnamefont {J.}~\bibnamefont
    {Kirschner}},\ }\bibfield  {title} {\bibinfo {title} {Decagonal
    quasiferromagnetic microstructure on the penrose tiling},\ }\href
    {https://doi.org/10.1103/PhysRevLett.90.137203} {\bibfield  {journal}
    {\bibinfo  {journal} {Phys. Rev. Lett.}\ }\textbf {\bibinfo {volume} {90}},\
    \bibinfo {pages} {137203} (\bibinfo {year} {2003})}\BibitemShut {NoStop}%
  \bibitem [{\citenamefont {Mellado}\ \emph {et~al.}(2012)\citenamefont
    {Mellado}, \citenamefont {Concha},\ and\ \citenamefont
    {Mahadevan}}]{Mellado2012}%
    \BibitemOpen
    \bibfield  {author} {\bibinfo {author} {\bibfnamefont {P.}~\bibnamefont
    {Mellado}}, \bibinfo {author} {\bibfnamefont {A.}~\bibnamefont {Concha}},\
    and\ \bibinfo {author} {\bibfnamefont {L.}~\bibnamefont {Mahadevan}},\
    }\bibfield  {title} {\bibinfo {title} {Macroscopic magnetic frustration},\
    }\href {https://doi.org/10.1103/PhysRevLett.109.257203} {\bibfield  {journal}
    {\bibinfo  {journal} {Phys. Rev. Lett.}\ }\textbf {\bibinfo {volume} {109}},\
    \bibinfo {pages} {257203} (\bibinfo {year} {2012})}\BibitemShut {NoStop}%
  \bibitem [{\citenamefont {Velo}\ \emph {et~al.}(2020)\citenamefont {Velo},
    \citenamefont {Cecchi}, \citenamefont {B{\'e}ron},\ and\ \citenamefont
    {Pirota}}]{Velo2020}%
    \BibitemOpen
    \bibfield  {author} {\bibinfo {author} {\bibfnamefont {M.~F.}\ \bibnamefont
    {Velo}}, \bibinfo {author} {\bibfnamefont {B.~M.}\ \bibnamefont {Cecchi}},
    \bibinfo {author} {\bibfnamefont {F.}~\bibnamefont {B{\'e}ron}},\ and\
    \bibinfo {author} {\bibfnamefont {K.~R.}\ \bibnamefont {Pirota}},\ }\bibfield
     {title} {\bibinfo {title} {Multipolar effects in the hysteresis behavior of
    {2D} arrays of compass needles: Experiments and simulations},\ }\href
    {https://doi.org/https://doi.org/10.1016/j.physo.2020.100041} {\bibfield
    {journal} {\bibinfo  {journal} {Phys. Open}\ }\textbf {\bibinfo {volume}
    {5}},\ \bibinfo {pages} {100041} (\bibinfo {year} {2020})}\BibitemShut
    {NoStop}%
  \bibitem [{\citenamefont {Gon\c{c}alves}\ \emph {et~al.}(2020)\citenamefont
    {Gon\c{c}alves}, \citenamefont {Gomes}, \citenamefont {Loreto}, \citenamefont
    {Nascimento}, \citenamefont {Moura-Melo}, \citenamefont {Pereira},\ and\
    \citenamefont {{de Araujo}}}]{Goncalves2020}%
    \BibitemOpen
    \bibfield  {author} {\bibinfo {author} {\bibfnamefont {R.}~\bibnamefont
    {Gon\c{c}alves}}, \bibinfo {author} {\bibfnamefont {A.}~\bibnamefont
    {Gomes}}, \bibinfo {author} {\bibfnamefont {R.}~\bibnamefont {Loreto}},
    \bibinfo {author} {\bibfnamefont {F.}~\bibnamefont {Nascimento}}, \bibinfo
    {author} {\bibfnamefont {W.}~\bibnamefont {Moura-Melo}}, \bibinfo {author}
    {\bibfnamefont {A.}~\bibnamefont {Pereira}},\ and\ \bibinfo {author}
    {\bibfnamefont {C.}~\bibnamefont {{de Araujo}}},\ }\bibfield  {title}
    {\bibinfo {title} {Naked-eye visualization of geometric frustration effects
    in macroscopic spin ices},\ }\href
    {https://doi.org/https://doi.org/10.1016/j.jmmm.2020.166471} {\bibfield
    {journal} {\bibinfo  {journal} {J. Mag. Mag. Mater}\ }\textbf {\bibinfo
    {volume} {502}},\ \bibinfo {pages} {166471} (\bibinfo {year}
    {2020})}\BibitemShut {NoStop}%
  \bibitem [{\citenamefont {Teixeira}\ \emph {et~al.}(2024)\citenamefont
    {Teixeira}, \citenamefont {Bernardo}, \citenamefont {Nascimento},
    \citenamefont {Saccone}, \citenamefont {Caravelli}, \citenamefont {Nisoli},\
    and\ \citenamefont {{de Araujo}}}]{Teixeira2024}%
    \BibitemOpen
    \bibfield  {author} {\bibinfo {author} {\bibfnamefont {H.}~\bibnamefont
    {Teixeira}}, \bibinfo {author} {\bibfnamefont {M.}~\bibnamefont {Bernardo}},
    \bibinfo {author} {\bibfnamefont {F.}~\bibnamefont {Nascimento}}, \bibinfo
    {author} {\bibfnamefont {M.}~\bibnamefont {Saccone}}, \bibinfo {author}
    {\bibfnamefont {F.}~\bibnamefont {Caravelli}}, \bibinfo {author}
    {\bibfnamefont {C.}~\bibnamefont {Nisoli}},\ and\ \bibinfo {author}
    {\bibfnamefont {C.}~\bibnamefont {{de Araujo}}},\ }\bibfield  {title}
    {\bibinfo {title} {Macroscopic magnetic monopoles in a {3D}-printed
    mechano-magnet},\ }\href
    {https://doi.org/https://doi.org/10.1016/j.jmmm.2024.171929} {\bibfield
    {journal} {\bibinfo  {journal} {J. Mag. Mag. Mater.}\ }\textbf {\bibinfo
    {volume} {596}},\ \bibinfo {pages} {171929} (\bibinfo {year}
    {2024})}\BibitemShut {NoStop}%
  \bibitem [{\citenamefont {Peroor}\ \emph {et~al.}(2025)\citenamefont {Peroor},
    \citenamefont {Scafuri}, \citenamefont {Bozhko},\ and\ \citenamefont
    {Iacocca}}]{Peroor2025}%
    \BibitemOpen
    \bibfield  {author} {\bibinfo {author} {\bibfnamefont {R.~R.}\ \bibnamefont
    {Peroor}}, \bibinfo {author} {\bibfnamefont {L.~A.}\ \bibnamefont {Scafuri}},
    \bibinfo {author} {\bibfnamefont {D.~A.}\ \bibnamefont {Bozhko}},\ and\
    \bibinfo {author} {\bibfnamefont {E.}~\bibnamefont {Iacocca}},\ }\bibfield
    {title} {\bibinfo {title} {Frequency comb in a macroscopic mechanomagnetic
    artificial spin ice},\ }\href
    {https://doi.org/10.1103/PhysRevApplied.23.044010} {\bibfield  {journal}
    {\bibinfo  {journal} {Phys. Rev. Appl.}\ }\textbf {\bibinfo {volume} {23}},\
    \bibinfo {pages} {044010} (\bibinfo {year} {2025})}\BibitemShut {NoStop}%
  \bibitem [{\citenamefont {Scafuri}\ \emph {et~al.}(2025)\citenamefont
    {Scafuri}, \citenamefont {Bozhko},\ and\ \citenamefont
    {Iacocca}}]{Scafuri2025}%
    \BibitemOpen
    \bibfield  {author} {\bibinfo {author} {\bibfnamefont {L.~A.}\ \bibnamefont
    {Scafuri}}, \bibinfo {author} {\bibfnamefont {D.~A.}\ \bibnamefont
    {Bozhko}},\ and\ \bibinfo {author} {\bibfnamefont {E.}~\bibnamefont
    {Iacocca}},\ }\bibfield  {title} {\bibinfo {title} {Wave dynamics in a
    macroscopic square artificial spin ice},\ }\href
    {https://doi.org/10.1103/PhysRevApplied.23.054093} {\bibfield  {journal}
    {\bibinfo  {journal} {Phys. Rev. Appl.}\ }\textbf {\bibinfo {volume} {23}},\
    \bibinfo {pages} {054093} (\bibinfo {year} {2025})}\BibitemShut {NoStop}%
  \bibitem [{\citenamefont {Chioar}\ \emph
    {et~al.}(2014{\natexlab{a}})\citenamefont {Chioar}, \citenamefont
    {Rougemaille}, \citenamefont {Grimm}, \citenamefont {Fruchart}, \citenamefont
    {Wagner}, \citenamefont {Hehn}, \citenamefont {Lacour}, \citenamefont
    {Montaigne},\ and\ \citenamefont {Canals}}]{Chioar2014}%
    \BibitemOpen
    \bibfield  {author} {\bibinfo {author} {\bibfnamefont {I.-A.}\ \bibnamefont
    {Chioar}}, \bibinfo {author} {\bibfnamefont {N.}~\bibnamefont {Rougemaille}},
    \bibinfo {author} {\bibfnamefont {A.}~\bibnamefont {Grimm}}, \bibinfo
    {author} {\bibfnamefont {O.}~\bibnamefont {Fruchart}}, \bibinfo {author}
    {\bibfnamefont {E.}~\bibnamefont {Wagner}}, \bibinfo {author} {\bibfnamefont
    {M.}~\bibnamefont {Hehn}}, \bibinfo {author} {\bibfnamefont {D.}~\bibnamefont
    {Lacour}}, \bibinfo {author} {\bibfnamefont {F.}~\bibnamefont {Montaigne}},\
    and\ \bibinfo {author} {\bibfnamefont {B.}~\bibnamefont {Canals}},\
    }\bibfield  {title} {\bibinfo {title} {Nonuniversality of artificial
    frustrated spin systems},\ }\href
    {https://doi.org/10.1103/PhysRevB.90.064411} {\bibfield  {journal} {\bibinfo
    {journal} {Phys. Rev. B}\ }\textbf {\bibinfo {volume} {90}},\ \bibinfo
    {pages} {064411} (\bibinfo {year} {2014}{\natexlab{a}})}\BibitemShut
    {NoStop}%
  \bibitem [{\citenamefont {Chioar}\ \emph
    {et~al.}(2014{\natexlab{b}})\citenamefont {Chioar}, \citenamefont {Canals},
    \citenamefont {Lacour}, \citenamefont {Hehn}, \citenamefont {Santos~Burgos},
    \citenamefont {Mente\ifmmode~\mbox{\c{s}}\else \c{s}\fi{}}, \citenamefont
    {Locatelli}, \citenamefont {Montaigne},\ and\ \citenamefont
    {Rougemaille}}]{Chioar2014b}%
    \BibitemOpen
    \bibfield  {author} {\bibinfo {author} {\bibfnamefont {I.~A.}\ \bibnamefont
    {Chioar}}, \bibinfo {author} {\bibfnamefont {B.}~\bibnamefont {Canals}},
    \bibinfo {author} {\bibfnamefont {D.}~\bibnamefont {Lacour}}, \bibinfo
    {author} {\bibfnamefont {M.}~\bibnamefont {Hehn}}, \bibinfo {author}
    {\bibfnamefont {B.}~\bibnamefont {Santos~Burgos}}, \bibinfo {author}
    {\bibfnamefont {T.~O.}\ \bibnamefont {Mente\ifmmode~\mbox{\c{s}}\else
    \c{s}\fi{}}}, \bibinfo {author} {\bibfnamefont {A.}~\bibnamefont
    {Locatelli}}, \bibinfo {author} {\bibfnamefont {F.}~\bibnamefont
    {Montaigne}},\ and\ \bibinfo {author} {\bibfnamefont {N.}~\bibnamefont
    {Rougemaille}},\ }\bibfield  {title} {\bibinfo {title} {Kinetic pathways to
    the magnetic charge crystal in artificial dipolar spin ice},\ }\href
    {https://doi.org/10.1103/PhysRevB.90.220407} {\bibfield  {journal} {\bibinfo
    {journal} {Phys. Rev. B}\ }\textbf {\bibinfo {volume} {90}},\ \bibinfo
    {pages} {220407} (\bibinfo {year} {2014}{\natexlab{b}})}\BibitemShut
    {NoStop}%
  \bibitem [{\citenamefont {Hofhuis}\ \emph {et~al.}(2020)\citenamefont
    {Hofhuis}, \citenamefont {Hrabec}, \citenamefont {Arava}, \citenamefont
    {Leo}, \citenamefont {Huang}, \citenamefont {Chopdekar}, \citenamefont
    {Parchenko}, \citenamefont {Kleibert}, \citenamefont {Koraltan},
    \citenamefont {Abert}, \citenamefont {Vogler}, \citenamefont {Suess},
    \citenamefont {Derlet},\ and\ \citenamefont {Heyderman}}]{Hofhuis2020}%
    \BibitemOpen
    \bibfield  {author} {\bibinfo {author} {\bibfnamefont {K.}~\bibnamefont
    {Hofhuis}}, \bibinfo {author} {\bibfnamefont {A.}~\bibnamefont {Hrabec}},
    \bibinfo {author} {\bibfnamefont {H.}~\bibnamefont {Arava}}, \bibinfo
    {author} {\bibfnamefont {N.}~\bibnamefont {Leo}}, \bibinfo {author}
    {\bibfnamefont {Y.-L.}\ \bibnamefont {Huang}}, \bibinfo {author}
    {\bibfnamefont {R.~V.}\ \bibnamefont {Chopdekar}}, \bibinfo {author}
    {\bibfnamefont {S.}~\bibnamefont {Parchenko}}, \bibinfo {author}
    {\bibfnamefont {A.}~\bibnamefont {Kleibert}}, \bibinfo {author}
    {\bibfnamefont {S.}~\bibnamefont {Koraltan}}, \bibinfo {author}
    {\bibfnamefont {C.}~\bibnamefont {Abert}}, \bibinfo {author} {\bibfnamefont
    {C.}~\bibnamefont {Vogler}}, \bibinfo {author} {\bibfnamefont
    {D.}~\bibnamefont {Suess}}, \bibinfo {author} {\bibfnamefont {P.~M.}\
    \bibnamefont {Derlet}},\ and\ \bibinfo {author} {\bibfnamefont {L.~J.}\
    \bibnamefont {Heyderman}},\ }\bibfield  {title} {\bibinfo {title} {Thermally
    superactive artificial kagome spin ice structures obtained with the
    interfacial {Dzyaloshinskii-Moriya} interaction},\ }\href
    {https://doi.org/10.1103/PhysRevB.102.180405} {\bibfield  {journal} {\bibinfo
     {journal} {Phys. Rev. B}\ }\textbf {\bibinfo {volume} {102}},\ \bibinfo
    {pages} {180405} (\bibinfo {year} {2020})}\BibitemShut {NoStop}%
  \bibitem [{\citenamefont {Wang}\ \emph {et~al.}(2007)\citenamefont {Wang}, ,
    \citenamefont {Li}, \citenamefont {McConville}, \citenamefont {Nisoli},
    \citenamefont {Ke}, \citenamefont {Freeland}, \citenamefont {Rose},
    \citenamefont {Grimsditch}, \citenamefont {Lammert}, \citenamefont {Crespi}
    \emph {et~al.}}]{Wang2007}%
    \BibitemOpen
    \bibfield  {author} {\bibinfo {author} {\bibfnamefont {R.}~\bibnamefont
    {Wang}}, , \bibinfo {author} {\bibfnamefont {J.}~\bibnamefont {Li}}, \bibinfo
    {author} {\bibfnamefont {W.}~\bibnamefont {McConville}}, \bibinfo {author}
    {\bibfnamefont {C.}~\bibnamefont {Nisoli}}, \bibinfo {author} {\bibfnamefont
    {X.}~\bibnamefont {Ke}}, \bibinfo {author} {\bibfnamefont {J.}~\bibnamefont
    {Freeland}}, \bibinfo {author} {\bibfnamefont {V.}~\bibnamefont {Rose}},
    \bibinfo {author} {\bibfnamefont {M.}~\bibnamefont {Grimsditch}}, \bibinfo
    {author} {\bibfnamefont {P.}~\bibnamefont {Lammert}}, \bibinfo {author}
    {\bibfnamefont {V.}~\bibnamefont {Crespi}}, \emph {et~al.},\ }\bibfield
    {title} {\bibinfo {title} {Demagnetization protocols for frustrated
    interacting nanomagnet arrays},\ }\href {https://doi.org/10.1063/1.2712528}
    {\bibfield  {journal} {\bibinfo  {journal} {J. Appl. Phys.}\ }\textbf
    {\bibinfo {volume} {101}},\ \bibinfo {pages} {09J104} (\bibinfo {year}
    {2007})}\BibitemShut {NoStop}%
  \bibitem [{\citenamefont {Nisoli}\ \emph {et~al.}(2007)\citenamefont {Nisoli},
    \citenamefont {Wang}, \citenamefont {Li}, \citenamefont {McConville},
    \citenamefont {Lammert}, \citenamefont {Schiffer},\ and\ \citenamefont
    {Crespi}}]{Nisoli2007}%
    \BibitemOpen
    \bibfield  {author} {\bibinfo {author} {\bibfnamefont {C.}~\bibnamefont
    {Nisoli}}, \bibinfo {author} {\bibfnamefont {R.}~\bibnamefont {Wang}},
    \bibinfo {author} {\bibfnamefont {J.}~\bibnamefont {Li}}, \bibinfo {author}
    {\bibfnamefont {W.~F.}\ \bibnamefont {McConville}}, \bibinfo {author}
    {\bibfnamefont {P.~E.}\ \bibnamefont {Lammert}}, \bibinfo {author}
    {\bibfnamefont {P.}~\bibnamefont {Schiffer}},\ and\ \bibinfo {author}
    {\bibfnamefont {V.~H.}\ \bibnamefont {Crespi}},\ }\bibfield  {title}
    {\bibinfo {title} {Ground state lost but degeneracy found: The effective
    thermodynamics of artificial spin ice},\ }\href
    {https://doi.org/10.1103/PhysRevLett.98.217203} {\bibfield  {journal}
    {\bibinfo  {journal} {Phys. Rev. Lett.}\ }\textbf {\bibinfo {volume} {98}},\
    \bibinfo {pages} {217203} (\bibinfo {year} {2007})}\BibitemShut {NoStop}%
  \bibitem [{\citenamefont {Ke}\ \emph {et~al.}(2008)\citenamefont {Ke},
    \citenamefont {Li}, \citenamefont {Nisoli}, \citenamefont {Lammert},
    \citenamefont {McConville}, \citenamefont {Wang}, \citenamefont {Crespi},\
    and\ \citenamefont {Schiffer}}]{Ke2008}%
    \BibitemOpen
    \bibfield  {author} {\bibinfo {author} {\bibfnamefont {X.}~\bibnamefont
    {Ke}}, \bibinfo {author} {\bibfnamefont {J.}~\bibnamefont {Li}}, \bibinfo
    {author} {\bibfnamefont {C.}~\bibnamefont {Nisoli}}, \bibinfo {author}
    {\bibfnamefont {P.~E.}\ \bibnamefont {Lammert}}, \bibinfo {author}
    {\bibfnamefont {W.}~\bibnamefont {McConville}}, \bibinfo {author}
    {\bibfnamefont {R.}~\bibnamefont {Wang}}, \bibinfo {author} {\bibfnamefont
    {V.~H.}\ \bibnamefont {Crespi}},\ and\ \bibinfo {author} {\bibfnamefont
    {P.}~\bibnamefont {Schiffer}},\ }\bibfield  {title} {\bibinfo {title} {Energy
    minimization and ac demagnetization in a nanomagnet array},\ }\href
    {https://doi.org/10.1103/PhysRevLett.101.037205} {\bibfield  {journal}
    {\bibinfo  {journal} {Phys. Rev. Lett.}\ }\textbf {\bibinfo {volume} {101}},\
    \bibinfo {pages} {037205} (\bibinfo {year} {2008})}\BibitemShut {NoStop}%
  \bibitem [{\citenamefont {Morgan}\ \emph {et~al.}(2013)\citenamefont {Morgan},
    \citenamefont {Bellew}, \citenamefont {Stein}, \citenamefont {Langridge},\
    and\ \citenamefont {Marrows}}]{Morgan2013}%
    \BibitemOpen
    \bibfield  {author} {\bibinfo {author} {\bibfnamefont {J.~P.}\ \bibnamefont
    {Morgan}}, \bibinfo {author} {\bibfnamefont {A.}~\bibnamefont {Bellew}},
    \bibinfo {author} {\bibfnamefont {A.}~\bibnamefont {Stein}}, \bibinfo
    {author} {\bibfnamefont {S.}~\bibnamefont {Langridge}},\ and\ \bibinfo
    {author} {\bibfnamefont {C.~H.}\ \bibnamefont {Marrows}},\ }\bibfield
    {title} {\bibinfo {title} {Linear field demagnetization of artificial
    magnetic square ice},\ }\href {https://doi.org/10.3389/fphy.2013.00028}
    {\bibfield  {journal} {\bibinfo  {journal} {Frontiers Phys.}\ }\textbf
    {\bibinfo {volume} {1}},\ \bibinfo {pages} {28} (\bibinfo {year}
    {2013})}\BibitemShut {NoStop}%
  \bibitem [{\citenamefont {Lammert}\ \emph {et~al.}(2010)\citenamefont
    {Lammert}, \citenamefont {Ke}, \citenamefont {Li}, \citenamefont {Nisoli},
    \citenamefont {Garand}, \citenamefont {Crespi},\ and\ \citenamefont
    {Schiffer}}]{Lammert2010}%
    \BibitemOpen
    \bibfield  {author} {\bibinfo {author} {\bibfnamefont {P.~E.}\ \bibnamefont
    {Lammert}}, \bibinfo {author} {\bibfnamefont {X.}~\bibnamefont {Ke}},
    \bibinfo {author} {\bibfnamefont {J.}~\bibnamefont {Li}}, \bibinfo {author}
    {\bibfnamefont {C.}~\bibnamefont {Nisoli}}, \bibinfo {author} {\bibfnamefont
    {D.~M.}\ \bibnamefont {Garand}}, \bibinfo {author} {\bibfnamefont {V.~H.}\
    \bibnamefont {Crespi}},\ and\ \bibinfo {author} {\bibfnamefont
    {P.}~\bibnamefont {Schiffer}},\ }\bibfield  {title} {\bibinfo {title} {Direct
    entropy determination and application to artificial spin ice},\ }\href
    {https://doi.org/10.1038/nphys1728} {\bibfield  {journal} {\bibinfo
    {journal} {Nat. Phys.}\ }\textbf {\bibinfo {volume} {6}},\ \bibinfo {pages}
    {786} (\bibinfo {year} {2010})}\BibitemShut {NoStop}%
  \bibitem [{\citenamefont {Rodr{\'\i}guez-Gallo}\ \emph
    {et~al.}(2021)\citenamefont {Rodr{\'\i}guez-Gallo}, \citenamefont
    {Ortiz-Ambriz},\ and\ \citenamefont {Tierno}}]{Rodriguez2021}%
    \BibitemOpen
    \bibfield  {author} {\bibinfo {author} {\bibfnamefont {C.}~\bibnamefont
    {Rodr{\'\i}guez-Gallo}}, \bibinfo {author} {\bibfnamefont {A.}~\bibnamefont
    {Ortiz-Ambriz}},\ and\ \bibinfo {author} {\bibfnamefont {P.}~\bibnamefont
    {Tierno}},\ }\bibfield  {title} {\bibinfo {title} {Topological boundary
    constraints in artificial colloidal ice},\ }\href
    {https://doi.org/10.1103/PhysRevLett.126.188001} {\bibfield  {journal}
    {\bibinfo  {journal} {Phys. Rev. Lett.}\ }\textbf {\bibinfo {volume} {126}},\
    \bibinfo {pages} {188001} (\bibinfo {year} {2021})}\BibitemShut {NoStop}%
  \bibitem [{\citenamefont {Baillou}\ \emph {et~al.}(2026)\citenamefont
    {Baillou}, \citenamefont {Terkel},\ and\ \citenamefont
    {Tierno}}]{Baillou2026}%
    \BibitemOpen
    \bibfield  {author} {\bibinfo {author} {\bibfnamefont {R.}~\bibnamefont
    {Baillou}}, \bibinfo {author} {\bibfnamefont {M.}~\bibnamefont {Terkel}},\
    and\ \bibinfo {author} {\bibfnamefont {P.}~\bibnamefont {Tierno}},\
    }\bibfield  {title} {\bibinfo {title} {Virtual magnetic hills to unlock the
    inner phases of hexagonal colloidal ice},\ }\href
    {https://doi.org/10.1039/D5SM01277E} {\bibfield  {journal} {\bibinfo
    {journal} {Soft Matter}\ }\textbf {\bibinfo {volume} {22}},\ \bibinfo {pages}
    {2122} (\bibinfo {year} {2026})}\BibitemShut {NoStop}%
  \bibitem [{\citenamefont {Nienhuis}\ \emph {et~al.}(1984)\citenamefont
    {Nienhuis}, \citenamefont {Hilhorst},\ and\ \citenamefont
    {Blote}}]{Nienhuis1984}%
    \BibitemOpen
    \bibfield  {author} {\bibinfo {author} {\bibfnamefont {B.}~\bibnamefont
    {Nienhuis}}, \bibinfo {author} {\bibfnamefont {H.~J.}\ \bibnamefont
    {Hilhorst}},\ and\ \bibinfo {author} {\bibfnamefont {H.}~\bibnamefont
    {Blote}},\ }\bibfield  {title} {\bibinfo {title} {Triangular {SOS} models and
    cubic-crystal shapes},\ }\href {https://doi.org/10.1088/0305-4470/17/18/025}
    {\bibfield  {journal} {\bibinfo  {journal} {J. Phys. A}\ }\textbf {\bibinfo
    {volume} {17}},\ \bibinfo {pages} {3559} (\bibinfo {year}
    {1984})}\BibitemShut {NoStop}%
  \bibitem [{\citenamefont {Millane}\ and\ \citenamefont
    {Blakeley}(2004)}]{Millane2004}%
    \BibitemOpen
    \bibfield  {author} {\bibinfo {author} {\bibfnamefont {R.~P.}\ \bibnamefont
    {Millane}}\ and\ \bibinfo {author} {\bibfnamefont {N.~D.}\ \bibnamefont
    {Blakeley}},\ }\bibfield  {title} {\bibinfo {title} {Boundary conditions and
    variable ground state entropy for the antiferromagnetic {Ising} model on a
    triangular lattice},\ }\href {https://doi.org/10.1103/PhysRevE.70.057101}
    {\bibfield  {journal} {\bibinfo  {journal} {Phys. Rev. E}\ }\textbf {\bibinfo
    {volume} {70}},\ \bibinfo {pages} {057101} (\bibinfo {year}
    {2004})}\BibitemShut {NoStop}%
  \bibitem [{\citenamefont {Smerald}\ and\ \citenamefont
    {Mila}(2018)}]{Smerald2018}%
    \BibitemOpen
    \bibfield  {author} {\bibinfo {author} {\bibfnamefont {A.}~\bibnamefont
    {Smerald}}\ and\ \bibinfo {author} {\bibfnamefont {F.}~\bibnamefont {Mila}},\
    }\bibfield  {title} {\bibinfo {title} {{Spin-liquid behaviour and the
    interplay between {Pokrovsky-Talapov} and {Ising} criticality in the
    distorted, triangular-lattice, dipolar {Ising} antiferromagnet}},\ }\href
    {https://doi.org/10.21468/SciPostPhys.5.3.030} {\bibfield  {journal}
    {\bibinfo  {journal} {SciPost Phys.}\ }\textbf {\bibinfo {volume} {5}},\
    \bibinfo {pages} {030} (\bibinfo {year} {2018})}\BibitemShut {NoStop}%
  \bibitem [{\citenamefont {Kasteleyn}(1963)}]{Kasteleyn1963}%
    \BibitemOpen
    \bibfield  {author} {\bibinfo {author} {\bibfnamefont {P.~W.}\ \bibnamefont
    {Kasteleyn}},\ }\bibfield  {title} {\bibinfo {title} {Dimer statistics and
    phase transitions},\ }\href {https://doi.org/10.1063/1.1703953} {\bibfield
    {journal} {\bibinfo  {journal} {J. Math. Phys.}\ }\textbf {\bibinfo {volume}
    {4}},\ \bibinfo {pages} {287} (\bibinfo {year} {1963})}\BibitemShut {NoStop}%
  \bibitem [{\citenamefont {Hamp}\ \emph {et~al.}(2018)\citenamefont {Hamp},
    \citenamefont {Moessner},\ and\ \citenamefont {Castelnovo}}]{Hamp2018}%
    \BibitemOpen
    \bibfield  {author} {\bibinfo {author} {\bibfnamefont {J.}~\bibnamefont
    {Hamp}}, \bibinfo {author} {\bibfnamefont {R.}~\bibnamefont {Moessner}},\
    and\ \bibinfo {author} {\bibfnamefont {C.}~\bibnamefont {Castelnovo}},\
    }\bibfield  {title} {\bibinfo {title} {Supercooling and fragile glassiness in
    a dipolar kagome {Ising} magnet},\ }\href
    {https://doi.org/10.1103/PhysRevB.98.144439} {\bibfield  {journal} {\bibinfo
    {journal} {Phys. Rev. B}\ }\textbf {\bibinfo {volume} {98}},\ \bibinfo
    {pages} {144439} (\bibinfo {year} {2018})}\BibitemShut {NoStop}%
  \bibitem [{\citenamefont {Cugliandolo}\ \emph {et~al.}(2020)\citenamefont
    {Cugliandolo}, \citenamefont {Foini},\ and\ \citenamefont
    {Tarzia}}]{Cugliandolo2020}%
    \BibitemOpen
    \bibfield  {author} {\bibinfo {author} {\bibfnamefont {L.~F.}\ \bibnamefont
    {Cugliandolo}}, \bibinfo {author} {\bibfnamefont {L.}~\bibnamefont {Foini}},\
    and\ \bibinfo {author} {\bibfnamefont {M.}~\bibnamefont {Tarzia}},\
    }\bibfield  {title} {\bibinfo {title} {Mean-field phase diagram and
    spin-glass phase of the dipolar kagome {Ising} antiferromagnet},\ }\href
    {https://doi.org/10.1103/PhysRevB.101.144413} {\bibfield  {journal} {\bibinfo
     {journal} {Phys. Rev. B}\ }\textbf {\bibinfo {volume} {101}},\ \bibinfo
    {pages} {144413} (\bibinfo {year} {2020})}\BibitemShut {NoStop}%
  \bibitem [{\citenamefont {Fisher}\ and\ \citenamefont
    {Selke}(1980)}]{Fisher1980}%
    \BibitemOpen
    \bibfield  {author} {\bibinfo {author} {\bibfnamefont {M.~E.}\ \bibnamefont
    {Fisher}}\ and\ \bibinfo {author} {\bibfnamefont {W.}~\bibnamefont {Selke}},\
    }\bibfield  {title} {\bibinfo {title} {Infinitely many commensurate phases in
    a simple {Ising} model},\ }\href
    {https://doi.org/10.1103/PhysRevLett.44.1502} {\bibfield  {journal} {\bibinfo
     {journal} {Phys. Rev. Lett.}\ }\textbf {\bibinfo {volume} {44}},\ \bibinfo
    {pages} {1502} (\bibinfo {year} {1980})}\BibitemShut {NoStop}%
  \bibitem [{\citenamefont {Rufino}\ \emph {et~al.}(2026)\citenamefont {Rufino},
    \citenamefont {Nyckees}, \citenamefont {Colbois},\ and\ \citenamefont
    {Mila}}]{Rufino2025}%
    \BibitemOpen
    \bibfield  {author} {\bibinfo {author} {\bibfnamefont {A.}~\bibnamefont
    {Rufino}}, \bibinfo {author} {\bibfnamefont {S.}~\bibnamefont {Nyckees}},
    \bibinfo {author} {\bibfnamefont {J.}~\bibnamefont {Colbois}},\ and\ \bibinfo
    {author} {\bibfnamefont {F.}~\bibnamefont {Mila}},\ }\bibfield  {title}
    {\bibinfo {title} {Topological {Devil}'s staircase in a constrained kagome
    {Ising} antiferromagnet},\ }\href {https://doi.org/10.1103/dgyr-5y75}
    {\bibfield  {journal} {\bibinfo  {journal} {Phys. Rev. Lett.}\ }\textbf
    {\bibinfo {volume} {136}},\ \bibinfo {pages} {086701} (\bibinfo {year}
    {2026})}\BibitemShut {NoStop}%
  \bibitem [{\citenamefont {Yokoi}\ \emph {et~al.}(1986)\citenamefont {Yokoi},
    \citenamefont {Nagle},\ and\ \citenamefont {Salinas}}]{Yokoi1986}%
    \BibitemOpen
    \bibfield  {author} {\bibinfo {author} {\bibfnamefont {C.~S.}\ \bibnamefont
    {Yokoi}}, \bibinfo {author} {\bibfnamefont {J.~F.}\ \bibnamefont {Nagle}},\
    and\ \bibinfo {author} {\bibfnamefont {S.~R.}\ \bibnamefont {Salinas}},\
    }\bibfield  {title} {\bibinfo {title} {Dimer pair correlations on the brick
    lattice},\ }\href {https://doi.org/10.1007/BF01011905} {\bibfield  {journal}
    {\bibinfo  {journal} {J. Stat. Phys.}\ }\textbf {\bibinfo {volume} {44}},\
    \bibinfo {pages} {729} (\bibinfo {year} {1986})}\BibitemShut {NoStop}%
  \bibitem [{\citenamefont {Jiang}\ and\ \citenamefont {Emig}(2006)}]{Jiang2006}%
    \BibitemOpen
    \bibfield  {author} {\bibinfo {author} {\bibfnamefont {Y.}~\bibnamefont
    {Jiang}}\ and\ \bibinfo {author} {\bibfnamefont {T.}~\bibnamefont {Emig}},\
    }\bibfield  {title} {\bibinfo {title} {Ordering of geometrically frustrated
    classical and quantum triangular {Ising} magnets},\ }\href
    {https://doi.org/10.1103/PhysRevB.73.104452} {\bibfield  {journal} {\bibinfo
    {journal} {Phys. Rev. B}\ }\textbf {\bibinfo {volume} {73}},\ \bibinfo
    {pages} {104452} (\bibinfo {year} {2006})}\BibitemShut {NoStop}%
  \bibitem [{\citenamefont {Kenyon}(2009)}]{Kenyon2009}%
    \BibitemOpen
    \bibfield  {author} {\bibinfo {author} {\bibfnamefont {R.}~\bibnamefont
    {Kenyon}},\ }\href {https://arxiv.org/abs/0910.3129} {\bibinfo {title}
    {Lectures on dimers}} (\bibinfo {year} {2009}),\ \Eprint
    {https://arxiv.org/abs/0910.3129} {arXiv:0910.3129} \BibitemShut {NoStop}%
  \bibitem [{\citenamefont {Samuel}(1980)}]{Samuel1980}%
    \BibitemOpen
    \bibfield  {author} {\bibinfo {author} {\bibfnamefont {S.}~\bibnamefont
    {Samuel}},\ }\bibfield  {title} {\bibinfo {title} {The use of anticommuting
    variable integrals in statistical mechanics. {I}. {T}he computation of
    partition functions},\ }\href {https://doi.org/10.1063/1.524404} {\bibfield
    {journal} {\bibinfo  {journal} {J. Math. Phys.}\ }\textbf {\bibinfo {volume}
    {21}},\ \bibinfo {pages} {2806} (\bibinfo {year} {1980})}\BibitemShut
    {NoStop}%
  \end{thebibliography}

%


\bigskip

\section*{Acknowledgements}

We thank Ga\"{e}tan Debontride and Thibault Ricart for valuable discussions and suggestions concerning the use of the mechanical shaker, Nicolas Roch for providing a useful piece of material, and Afonso Rufino for discussions around Ising-to-fermionic mappings.

\section*{Funding Statement}

The work was supported by Agence Nationale de la Recherche through projects No. ANR-22-CE30-0041-01 `ArtMat' and ANR-25-CE30-5029 `MetaCharge'.

\section*{Author Contributions Statement}

R.D. designed and implemented the experimental setup, performed the measurements, developed the tools for automated analysis of the experimental data, wrote Monte Carlo codes, interpreted the results and wrote the first draft of the manuscript. L.D.R. fabricated the lattices. J. Colbois contributed in the interpretation of the results. J. Coraux wrote Monte Carlo codes and contributed to the analysis of the data. N.R. and J. Coraux contributed to the interpretation of the results, supervised the project and prepared the final version of the manuscript. All authors edited and revised the manuscript.

\section*{Competing Interests Statement}

The authors declare no competing interests.

\section*{Figure Legends/Captions}

\noindent
\textbf{Figure 1:} Macroscopic TIAF emulator. (a) Schematics of the triangular lattice of cavities, filled with NdFeB cylinders, and driven in sinusoidal motion $z(t)$ by a mechanical shaker. Lattices with $N=7$ and 11 magnets per edge have been manufactured. (b) Three neighbour magnets; $D=8.7$~mm ($N=7$) or $D=6.0$~mm ($N=11$) is the lattice period and $\zeta \in[-\Delta/2,+\Delta/2]$, the magnet position, with $\Delta=3$~mm, which defines the $\pm$1 value of an Ising degree of freedom. (c) Interaction potential $U$ vs $\zeta$; an energy barrier $E_\mathrm{b}$ separates the two degenerate low-energy states.

\bigskip

\noindent
\textbf{Figure 2:} Naked eye visualisation of equilibration and concerted spin dynamics. (a) Evolution of the first five spin correlators $C_{k\leq5}$ during a stepwise demagnetization, starting from an almost saturated configuration, and as $\Delta z$ increases stepwise over time (step duration: 0.1~s). (b) $C^{\mathrm{MC}}_{k\leq5}$ as function of temperature from Monte Carlo simulations (shaded areas represent the standard deviation), experimental data points at 60~s (colored dots) and cumulative correlator residual $r(T)$. (c) Experimental configuration obtained at 60~s, with all bulk $\threeUp$ / $\threeDown$ excitations, and a few of the edge $\twoUp$ / $\twoDown$ excitations highlighted. (d) $\chi(T)$ as function of time. The temperature minimising $\chi$ is overlaid, and becomes an equilibrium temperature when $\chi<2$.

\bigskip

\noindent
\textbf{Figure 3:} Low and tunable equilibrium temperatures. (a) Cumulative residual of the five first correlators relative to Monte Carlo simulations (OBC), as function of temperature, using a driving protocol. The 40 experimental configurations have been grouped in nine collections of configurations, and their average $C_{k\leq5}$ correlators each produce a curve. Curves have been vertically shifted, each by 0.1, for clarity. The curves' minima define an effective equilibrium temperature. The configuration shown as inset corresponds to the lowest (effective) equilibrium temperature. (b) Magnetic structure factors at different (effective) temperatures, from the Monte Carlo and experimental data. (c) Summary of effective equilibrium temperatures (standard deviations are smaller than the symbol size) for lattices of sizes $N=7$ and $N=11$ using the driving protocol, varying the quench time and agitation frequency (for $N=11$); the effective equilibrium temperature reached for $N=7$ via stepwise demagnetisation (as in Fig.~\ref{fig:dynamics}b), and the value obtained with arrays of nanomagnets \cite{Pip2021,Pac2025} (red lines) are also shown. X$_7$ and X$_{11}$ mark the crossover temperatures for the two lattice sizes. (d) Conditional entropy (with its experimental standard deviation) corresponding to the temperatures in (c). Expectations for the TIAF's (PBC, $N=11$) ground state and a randomly disordered configuration are shown as horizontal lines.

\bigskip

\noindent
\textbf{Figure 4:} Thermodynamics of the TIAF under different boundary conditions. (a) Entropy density ($s$) and (b) specific heat ($c_\mathrm{v}$, together with its standard deviation shown as a shaded area) as function of temperature for a rhombus-shaped TIAF in periodic boundary conditions (rPBC, 121 spins) and hexagon-shaped lattices (two sizes, $N$ = 7, 11, so 127 and 331 spins) in open boundary conditions (hOBC), according to Monte Carlo simulations (see Methods). 

\bigskip

\noindent
\textbf{Figure 5:} Boundary condition engineering and ground state topological sectors. (a-c) Experimental configurations with fixed (a) alternated, (b) disordered, (c) mixed alternated+all-down and (d) mixed alternated+disorder arrangements of spins at the lattice edges (along the orange lines). All configurations comply with PBC. Note the absence of excitations in (a,b). Below each configuration, MSF averaged over 32 (a,b), 20 (c) and 40 (d) different configurations is shown. (e) Triangular lattice's dual, a honeycomb lattice (pink). (f) Dimer mapping for the configuration in (b), with a green dimer placed on the segments of the dual lattice that are surrounded by $\twoUp$ or $\twoDown$ doublets. Dimers away from the lattice edges are not represented. Each dimer is associated to a winding, with positive or negative sign depending on the lattice edge, and the winding numbers are sums over two paths, defining the topological numbers $W_1$ and $W_2$. (g) Allowed pairs (red points) of topological winding numbers ($W_1$,$W_2$) in the TIAF's ground state (in PBC). The configurations shown in (a,b) both correspond to the (0,0) topological sector. The configurations in (c,d) ressemble configurations in the stripe (-6,0) and (4,4) sectors respectively; however they comprise $\threeUp$ and $\threeDown$ excitations and thus do not belong to the ground state. (h) Conditional entropy (and its experimental standard deviation) in the (a-d) kinds of configurations as function of the number of excitation in the configurations, compared to the expectations for the (0,0) ground state (horizontal lines, corresponding to edge configurations as in (a), in brown, and in (b), in blue).

\end{document}